\documentclass[referee,usenatbib]{mn2e}
\usepackage{graphicx,amssymb,natbib}

\title{Stellar contents and star formation in the young star cluster Be 59}

\author[Pandey et al.]
       {A. K. Pandey$^{1,2,3}$ \thanks{pandey@aries.ernet.in}, Saurabh Sharma$^{3}$, K. Ogura$^{4}$, D. K. Ojha$^2$, 
\newauthor W. P. Chen$^1$, B. C. Bhatt$^{5}$ and S. K. Ghosh$^2$,\\\\
$^1$Institute of Astronomy, National Central University, Chung-Li 32054, Taiwan\\
$^2$Tata Institute of Fundamental Research, Mumbai - 400 005, India\\
$^3$Aryabhatta Research Institute of Observational Sciences (ARIES), Manora Peak, Nainital, 263 129, India\\
$^4$Kokugakuin University, Higashi, Shibuya-ku, Tokyo 150-8440, Japan\\
$^5$CREST, Indian Institute of Astrophysics, Hosakote 562 114, India}
\date{Accepted ......
      Received .....}

\pagerange{\pageref{firstpage}--\pageref{lastpage}}
\pubyear{2007}

\begin{document}

\maketitle

\label{firstpage}

\begin{abstract}

We present $UBVI_C$ CCD photometry of the young open cluster Be 59 with the aim to study 
the star formation scenario in the cluster. The radial extent of the cluster is found to be
$\sim$ 10 arcmin (2.9 pc). The interstellar extinction in the cluster region varies between
 $E(B-V) \simeq$ 1.4 to 1.8 mag. The ratio of total-to-selective extinction in the cluster region
is estimated as $3.7\pm0.3$. The distance of the cluster is found to be $1.00\pm0.05$ kpc.
Using near-infrared colours and slitless spectroscopy, we have identified young stellar objects (YSOs) 
in the open cluster Be 59 region. The ages of these YSOs range between $<1$ Myr to $\sim$ 2 Myr, 
whereas the mean age of the massive stars in the cluster region is found to be $\sim$ 2 Myr. There is evidence for second generation star formation outside the boundary of the cluster, which may be triggered by massive stars in the cluster. The slope of the initial mass function, $\Gamma$,
in the mass range $2.5 < M/M_\odot \le 28$ is found to be $-1.01\pm0.11$ which is shallower than the Salpeter value (-1.35), whereas in the mass range $1.5 < M/M_\odot \le 2.5$ the slope is almost flat. The slope of the K-band luminosity function is estimated as $0.27\pm0.02$, which is smaller than the average value ($\sim$0.4) reported for young embedded clusters. 
Approximately $32\%$ of H$\alpha$ emission stars of Be 59 exhibit NIR excess indicating that inner disks 
of the T-Tauri star (TTS) population have not dissipated. 
The MSX and IRAS-HIRES images around the cluster region are also used to study the emission from unidentified
infrared bands and to estimate the spatial distribution of optical depth of warm and cold interstellar
dust.

\end{abstract}


\section{Introduction}

High-mass star forming regions have been known for many years as OB associations and HII regions and they  have been observed quite extensively on various aspects. However, the census of low mass stars in such regions has not been possible until recently. Recent advancement in detectors have permitted the detection of substantial population of low mass stars in OB associations. Recently relatively large number of low mass stars have been detected in a few OB associations (e.g. Upper Scorpios, the $\sigma$ and $\lambda$ Ori regions; Preibisch \& Zinnecker 1999, Dolan \& Mathieu 2002). Since this realization, surveys have demonstrated that the Initial Mass Functions (IMFs) are essentially the same in all star forming regions. The apparent difference may be due mainly to the  inherent low percentage of high mass stars and the incomplete survey of low mass stars in high-mass star forming regions (e.g. Kroupa 2002, Preibisch \& Zinneker 1999, Hillenbrand 1997, Massey et al. 1995).

The massive stars of OB associations have strong influence and significantly affect the entire star-forming  region. Strong UV radiation from massive stars could evaporate nearby clouds and consequently terminate star formation. Herbig (1962) suggested that low and intermediate-mass stars form first and with the formation of most massive star in the region, the cloud gets disrupted and star formation ceases.
Alternatively shock waves associated with the ionization front may squeeze molecular cloud and induce subsequent star formation. Elmegreen \& Lada (1977) proposed the expanding ionization fronts play a constructive role to incite a sequence of star-formation activities in the neighborhood. The question however is: could both of these effects occur in different parts of the same star-forming regions? 

Recent studies have shown cases where massive stars ionize adjacent clouds revealing embedded stars (e.g. the Eagle Nebula; Hester et al. 1996). Preibisch \& Zinnecker (1999) show that star formation in the upper Scorpius OB association was likely triggered by a nearby supernovae. Recently Lee et al. (2005) have found evidence for triggered star formation in the bright-rimmed clouds (BRCs) in the vicinity of O stars. Pandey et al. (2001, 2005) found that star formation in few young clusters may be a continuous process. For example, In the case of NGC 663, star formation seems to have taken place non-coevally in the sense that formation of low mass stars precedes the formation of most massive stars. Whereas, in the case of NGC 654 and NGC 3603, formation of low mass stars did not cease after the formation of most massive stars in the clusters.

The Sharpless region S171 is a large HII region associated with the Cepheus OB4 stellar association (Yang \& Fukui 1992). The star cluster Be 59 is located at the center of the nebulous region with nine  O7 - B3 stars. The physical properties of the ionized gas have been investigated in radio continuum and
recombination lines. (e.g. Felli et al. 1977, Rossano et al. 1980, Harten et al. 1981). 
Yang \& Fukui (1992) have mapped the molecular gas distribution using the $J=1-0$ lines 
of $^{12}$CO and $^{13}$CO emission, indicating that Be 59 generates the ionization front 
on the surface of two dense molecular clouds. They suggested that the dense gas is in 
contact with the continuum source and ionization front is deriving shocks into the clumps. 
As the shock penetrates into the molecular clumps, it forms a compressed gas layer which is 
unstable against self-gravity (Elmegreen 1989, 1998). One can expect that a new generation 
of stars may form from the compressed gas layer. Thus the S171 region provides an opportunity 
to make a comprehensive exploration of the effects of massive stars on low mass star formation. 
Therefore in this paper, we present a multi-wavelength study of the star forming region S171. 
We have also carried out slitless spectroscopy 
to identify pre-main sequence H$\alpha$ emission stars in the region. 
In section 2 we present optical CCD photometric and slitless spectroscopic observations 
and brief description of data reduction. 
In sect 3, we discuss archival data set used in the present study. In the ensuing sections we 
discuss results and star formation scenario in the Be 59 region.

\section{Observations and data reduction}

\subsection{$UBV{I}_c$ CCD observations}

The CCD $UBVI_c$ observations of the cluster were obtained using the 105-cm Schmidt telescope of the Kiso Observatory on November 19-21, 2001. The CCD camera used a SITe 2048 $\times$ 2048 TK2048E chip having a pixel size $24 \mu$m. At the Schmidt focus (f/3.1) each pixel corresponds to 1.5 arcsec and the entire chip covers a field of $\sim 50 \times 50$ arcmin$^2$ on the sky. The read-out noise and gain of the CCD are 23.2 $e^{-}$ and 3.4 $e^{-}/ADU$ respectively. 
A number of short and deep exposure frames were taken. The log of the observations is given in Table 1. 
A number of bias and dome flat frames were also taken during the observing runs. 
The observations were standardized by observing 14 standard stars in SA95 field (Landolt 1992) with brightness $12.2<V<15.6$ and colour $0.45<(B-V)<1.51$ on 2001 November 22.

The data analysis was carried out at Aryabhatta Research Institute of Observational
Sciences (ARIES), Nainital. Initial processing of the data was done using
IRAF\footnote{IRAF is distributed by National Optical Astronomy
Observatories, USA} and ESO-MIDAS\footnote{ESO-MIDAS is developed and
maintained by the  European Southern Observatory} 
data reduction software packages. The photometry was carried
out by using DAOPHOT-II software (Stetson 1987). The point spread function (PSF)
was obtained for each frame using several uncontaminated stars. FWHM of the star
images  was $\sim3^{\prime \prime}$. When bright stars were saturated on deep
exposure frames, their magnitudes were taken from short exposure frames. Calibration
of the instrumental to the standard system was done using the procedure outlined
by Stetson (1992). The details of the procedure,  zero-point constants, colour coefficients
and extinction coefficients are given elsewhere (Sharma et al. 2006).
Fig. 1 shows the standardization residuals, $\Delta$, between the standard and transformed
$V$ magnitudes and $(U-B),(B-V)$ and $(V-I)$ colours of standard stars.
The standard deviation in $\Delta V$, $\Delta (U-B)$, $\Delta (B-V)$ 
and $\Delta (V-I)$ are  0.010, 0.055, 0.023 and 0.023 mag respectively. 
The photometric errors increase with $V$ magnitude and becomes
larger ($\ge 0.1$ mag) for stars fainter than $V=19.5$ mag. The measured photometric data along
with positions of all the stars are given in Table 2\footnote{Table 2 is available in electronic form only},
a sample of which is given in this paper. 

To study the luminosity function  (LF) and mass function (MF) it is very important to make necessary corrections in the data sample to take into account the incompleteness that may occur for various reasons (e.g. due to the crowding of stars). We used the ADDSTAR routine of DAOPHOT II to determine the completeness factor (CF). The procedure has been outlined in detail in our earlier work (Pandey et al. 2001, 2005). The sample is complete at the brighter magnitude level ($V \le 14$). The incompleteness increases with increasing magnitude as expected (e.g. CF $\sim 0.84$ for $V=19$).

\subsection{Grism slitless spectroscopy}

The observations were also carried out in slitless mode with a grism as the dispersing element using the Himalayan Faint Object Spectrograph Camera (HFOSC) instrument at the 2-m Himalayan Chandra Telescope (HCT) on 22 September 2004 and 7 - 9 January 2005.  This mode of observation using the HFOSC yields an image where the stellar images are replaced by their spectra. A combination of a wide H$\alpha$ filter ($\lambda = 6300 - 6740 \rm \AA$) and Grism 5 (5200 - 10300 $ \rm \AA$) of HFOSC was used without any slit. The resolution of the grism is 870. The central $2K\times2K$ pixels of the $2K\times4K$ CCD 
were used for data acquisition. The pixel size is 15 micron with the scale of 0.297 arcsec/pixel.
We secured three spectroscopic frames of 7 min exposure with grism in and three direct frames of 
1 minute exposure with the grism out. The seeing during the observations was $\sim$ 1.5 to 2.3 arcsec. 
The co-added spectroscopic frames yield limiting magnitude 
of about $V\sim20$ for continuum detection. The slitless spectra thus obtained were used to identify
the emission line stars in the field of the cluster. The data were reduced following the procedures with IRAF as described by Ogura et al. (2002). The equivalent width (EW) has also been calculated as described by Ogura et al. (2002) and relevant information's about the detected $H\alpha$ stars in Be 59 are given in Table 3.
The frequency distribution of the H$\alpha$ EWs reveals that 23\% of H$\alpha$ stars 
have EW $\le 10 \rm \AA$ which is significantly less than the fraction (60\%) reported 
by Herbig (1998) in the case of IC 348. The possible cause for this deficiency is the poor detection efficiency of smaller EWs in the present survey due to seeing conditions.

\section{Archival datasets}

Near-Infrared (NIR) $JHK_s$ data for point sources around Be 59 have been obtained from the Two Micron All Sky Survey (2MASS) Point Source Catalog. The 2MASS database provides unbiased photometry in the $J$ $(1.25\mu$m), $H$ $(1.65 \mu$m) and $K_s$ $(2.17 \mu$m) bands to a limiting magnitude of 15.8, 15.1 and 14.3 respectively, with a signal to noise ratio (S/N) of 10. To improve the photometric accuracy, we use photometric quality flag (ph-qual=AAA) which gives a S/N$\ge10$ and photometric uncertainty $<$ 0.10 mag.

The Midcourse Space experiment (MSX) surveyed the Galactic plane within the range $|b|\le5^\circ$ in the four mid-infrared bands between 8 and 21 $\mu$m at a spatial resolution of $\sim18^{\prime \prime}.3$ (Price et al. 2001). Two of these bands (A and C) with ${\it \lambda(\Delta \lambda)}$ corresponding to 8.28(3.36) $\mu$m and 12.13(1.71) $\mu$m include several Unidentified Infrared emission Bands (UIBs) at 6.2, 7.7, 8.7 11.3, and 12.7 $\mu$m. MSX images in the four bands around the cluster region were used to study the emission from the UIBs and to estimate the spatial distribution of optical depth
of the warm interstellar dust.

The data from the IRAS survey in the four bands (12, 25, 60 and 100 $\mu$m) for the
region around Be 59 were HIRES  processed (Aumann et al. 1990) to obtain
high angular resolution maps. These maps have been used to study the spatial
distribution of optical depth of the cold interstellar dust. The IRAS point sources have 
also been identified in the cluster region and their details are given in Table 4.

\section{Results}

\subsection{Structure of the cluster}

\subsubsection{Isodensity contours}

Internal interaction of two-body relaxation due to encounters among members
stars and external tidal forces due to the Galactic disk or giant molecular
clouds can significantly influence the morphology of the clusters. However, in the
case of young clusters, where dynamical relaxation is not important because of cluster's
young age, the  stellar distribution can be considered as the initial morphology
of the cluster that should be governed by the initial conditions in the parent
molecular cloud (Chen et al. 2004). To study the morphology of the clusters
we plotted isodensity contours for the sample of main sequence (MS) (cf. Pandey et al. 2001) 
stars having $V\le 18$ mag (Fig. 2 {\it(left panel)}). Similarly the isodensity 
contours using the 2MASS data are also plotted in Fig. 2 {\it(right panel)}. 
Both NIR and optical data indicate an elongated morphology for the cluster.

\subsubsection{Radial stellar surface density profile}

To find out the extent of the cluster we used optical data of probable
MS stars ($V\le 16$ and $18$ mag) as well as 2MASS data ($K\lesssim14.3$).
The center of the cluster was determined using the stellar density
distribution in a 100 pixel wide strip along both X and
Y directions around an initially eye estimated center. The point of maximum density
obtained by fitting Gaussian distribution is considered as the center
of cluster. The X,Y pixel coordinates of the cluster for optical data
are found to be $1128\pm7, 1024\pm13$ respectively, which correspond to 
$\rm{RA}(2000) = 00^{\rm{h}} 02^{\rm{m}} 10^{\rm{s}}.4 \pm 0.7^{\rm{s}}$, $\rm{Dec}.~(2000) = 67^{\circ} 25^{'} 10^{''} \pm 19^{''}$.

To determine the radial surface density we divided the cluster into a number 
of concentric circles. Projected radial stellar density in each concentric annulus 
was obtained by dividing the number of stars by 
its area. Densities thus obtained are plotted in Fig. 3. The error bars are
derived assuming that the number of stars in a concentric annulus follows
the Poisson statistics. The horizontal dashed line in the plot indicates the
density of contaminating field stars, which is obtained from the region (about 1.5 cluster radius)
outside the cluster area.
The extent of the cluster $r_{cl}$ is defined as the point at which the
radial density becomes constant and merges with the field star density. The
radial extent of the cluster $r_{cl}$ from optical data and NIR 2MASS
data is estimated as $\sim10^ \prime$ (2.9 pc) for a distance of 1.0 kpc.

The observed radial density profile (RDP) of the clusters was parametrized following the
approach by Kaluzny \& Udalski (1992) where the projected radial density $\rho(r)$
is described as:
\begin{center}
$\rho (r) = {f_0\over \displaystyle{1+\left({r\over r_c}\right)^2}}$,
\end{center}
where the core radius $r_c$ is the radial distance at which the value of $\rho (r)$ becomes half of the central density $f_0$. The best fit obtained by $\chi^2$ minimization technique, is shown in Fig. 3. Within the uncertainties the model reproduces well the radial density profile of the cluster. The core radius $r_{c}$ increases from $1^\prime.0\pm0^\prime.1$ for $V\le16$ mag to $1^\prime.9 \pm0.^\prime3$ for $V\le18$ mag, whereas for the 2MASS data $r_c$ comes
out to be $1.^\prime4\pm0.^\prime2$. A closer look of the RDPs indicates
a lack of stars in optical data in the $5^\prime<r<8^\prime$ whereas there
seems an increase in the stellar density obtained from the 2MASS data in this radial distance range.
This enhancement may be due to embedded population of low mass stars.

\subsection{Interstellar extinction}

The interstellar extinction in the cluster region is studied using the $(U-B)/(B-V)$ 
two-colour diagram (TCD) shown in Fig. 4. In Fig. 4 the zero-age-main-sequence (ZAMS) from Schmidt-Kaler (1982) is shifted along a 
normal reddening vector having a slope of $E(U-B)/E(B-V) = 0.72$. A foreground population 
reddened by $E(B-V)$=0.4 mag can be noticed easily. The TCD shows a variable reddening in 
the cluster region between $E(B-V)_{min} \simeq 1.4$ mag and $E(B-V)_{max} \simeq 1.8$ mag.

The reddening for the individual stars (those with photometric $V_{error} \le 0.1$ mag) 
with spectral type earlier than A0 has also been derived using the Q method 
(Johnson \& Morgan 1953). The radial variation of reddening in the cluster 
region is shown in Fig. 5. The error bars represent standard errors. 
The reddening does not show any significant spatial variation in the cluster region barring a peak at $r\sim7.^\prime5$. It may interesting to point out that around the maxima of $E(B-V)$ at $r\sim7.^\prime5$, the optical RDP show decrease in stellar density. The decrease in stellar density becomes more prominent if we include fainter stars in the sample. Whereas, at $r\sim7^\prime.5$, the RDP from 2MASS data shows enhancement in the stellar density.

To study the nature of the extinction law in the cluster region, we used TCDs as 
described by Pandey et al. (2000, 2003). The TCDs of the form of ($V-\lambda$) vs. ($B-V$), 
where $\lambda$ is one of the wavelengths of the broadband filters $R,I,J,H,K$ and $L$ 
provide an effective method for separating the influence of the normal extinction 
produced by the diffuse interstellar medium from that of the abnormal extinction 
arising within regions having a peculiar distribution of dust sizes 
(cf. Chini \& Wargau 1990, Pandey et al. 2000). The TCDs for the cluster region are shown in Fig. 6. 
The slopes of the distributions, $m_{cluster}$, (cf. Pandey et al. 2003) is found to 
be $1.34\pm0.08, 2.30\pm0.08,2.90\pm0.11, 3.10\pm0.13$ for $(V-I), (V-J), (V-H), (V-K)$ vs. $(B-V)$ 
diagrams respectively. The ratios ${E(V-\lambda)}\over {E(B-V)}$ and the 
ratio of total-to-selective extinction in the cluster region, $R_{cluster}$, is derived using the procedure given by Pandey et al. (2003). The ratios ${E(V-\lambda)}\over {E(B-V)}$  $(\lambda \ge \lambda_I)$ yield $R_{cluster} = 3.7 \pm 0.3$ which indicates an anomalous reddening law.  In the central region of Be 59 MacConnell (1968) has also 
found evidence for large value (3.4-3.7) of $R$.
Several studies have already pointed anomalous reddening law with  high $R$ value in the vicinity of star forming regions (see e.g. Pandey et al. 2003 and references therein), however for the Galactic diffuse interstellar medium a normal value of $R=3.1$ is well accepted. The higher than the normal value of $R$ have been attributed to the presence of  larger dust grains. There is evidence that within dark clouds accretion of ice mantles on grains and coagulation due to colliding grains changes the size distribution towards larger particles. On the other hand, in star forming regions, radiation from massive stars may evaporate ice mantles resulting in small particles. Here it is interesting to mention that Okada et al. (2003), on the basis of the [Si II] 35 $\mu$m to [N II] 122 $\mu$m ratio, suggested that efficient dust destruction is undergoing in the ionized region. Chini \& Kr\"{u}gel (1983) and Chini \& Wargau (1990) have shown that both - larger and smaller grains - may increase the ratio of total-to-selective extinction.
 
\subsection{Optical Colour Magnitude Diagrams: Distance, Age and Probable members of the cluster}

Colour-magnitude-diagram (CMD) for stars lying within the cluster region are shown in Fig. 7.  The effect of field star contamination is apparent in the CMDs of the outer/whole cluster region. The field population, 
which is reddened by $E(B-V)$=0.4 mag is found to be located at $\sim$470 pc (See Fig. 7). 
Since foreground reddening ($E(B-V)_{min}$=1.4 mag) towards the cluster region is found to be normal, we used $E(U-B)/ E(B-V)$=0.72, $A_V=3.1 \times E(B-V)$ and $E(V-I)= 1.25 \times E(B-V)$ to visually fit the theoretical isochrone of 4 Myr (Z=0.02) by Bertelli et al. (1994) to the blue envelope of the observed MS of the cluster and found a distance modulus $(m-M)_V = 14.35\pm0.10$ which corresponds to a distance of $1.00\pm0.05$ kpc. The distance estimate is in agreement with that recently obtained by Sharma et al. (2007a).

To study the luminosity function (LF) and mass function (MF), it is necessary to remove field star contamination from the sample of stars in the cluster region. Membership determination is also crucial for assessing the presence of pre-main-sequence (PMS) stars, because PMS stars and dwarf foreground stars both occupy similar positions above the ZAMS in the CMD. In the absence of  proper motion study, we used statistical criterion to estimate the number of probable member stars in the cluster region.

Figs. 8a and 8b show $V, (B-V)$ CMDs for the cluster and field regions respectively. The contamination due to field population is clearly visible in the cluster region CMD. To remove contamination of field stars from the MS and PMS sample, we statistically subtracted the contribution of field stars from the CMD of the  cluster region using following procedure. For any  star in the $V, (B-V)$ CMD of the field region, the nearest star in the cluster's $V,(B-V)$ CMD within $V\pm0.125$ and $(B-V)\pm0.065$ of the field star was removed. While removing stars from the cluster CMD, necessary corrections for incompleteness of the data samples were taken into account.  The statistically cleaned $V, (B-V)$ CMD of the cluster region is shown in Fig. 8c which clearly shows the presence of PMS stars in the cluster. Fig. 9 shows statistically cleaned unreddened $V_0,(B-V)_0$ CMD where stars having spectral type earlier than $A0$ were individually dereddened (cf. \S 4.2), whereas the mean reddening, estimated from available individual reddening values in that region, was used for other stars. Since the cluster region shows anomalous reddening the intrinsic magnitudes of stars having spectral type earlier than $A0$ were obtained as follows;

$V_0 = V - (3.1\times E(B-V)_{min} + R_{cluster} \times \Delta E(B-V))$

where $\Delta E(B-V) = E(B-V)_* - E(B-V)_{min}$ and  $E(B-V)_*$ is $E(B-V)$ value of individual stars estimated from the $Q$ method. $E(B-V)_{min}$ = 1.4 mag represents reddening due to diffuse normal interstellar matter. 
Proper motions of 9 stars in the cluster region are available in the WEBDA (Mermilliod 1995). 
The membership probabilities along with star numbers from WEBDA are shown in Fig. 9. 
The photometric data for 2 stars (star number 3 and 4) are not available in present 
sample, so were taken from WEBDA. The O7 star (star number 4) in the region has a 
very low membership probability (p=0.08).

In Fig. 9 the ZAMS by Schmidt-Kaler (1982), isochrones for 4 Myr by Bertelli et al. (1994) and PMS isochrones by Siess et al. (2000) have also been plotted. Fig. 9 yields an average post-main-sequence age of massive stars of the cluster as $\sim$ 2 Myr and also indicates a significant spread in the age of PMS stars. To check the reality of the age spread of PMS population, we plotted  $V,(V-I)$ CMD for $H\alpha$ emission stars and NIR excess stars (see \S 4.4) in Fig. 10, which also indicates an age spread of about $\sim$1-5 Myr for the PMS stars.

\subsection {NIR colour-colour and colour-magnitude diagrams}

NIR imaging surveys have emerged as a powerful tool to detect low-mass stars in star forming regions. The $(J-H)/(H-K)$ colour-colour diagrams for the cluster region ($r<r_{cl}$) and for a nearby reference field are shown in Fig. 11. The solid and thick dashed curves represent the unreddened main-sequence and giant branch (Bessell \& Brett 1988) respectively. The parallel dashed lines are the reddening vectors for early and giant stars (drawn from the base and tip of the two branches). The dotted line indicates the locus of T Tauri stars (Meyer et al. 1997). The extinction ratios $A_J/A_V = 0.265, A_H/A_V = 0.155$ and $A_K/A_V=0.090$ have been taken from Cohen et al. (1981). 
All the 2MASS magnitudes and colours have been converted into the CIT system. The curves are also in the CIT system. We classified sources into three regions in the colour-colour diagram (cf. Ojha et al. 2004a). `F' sources are located between the reddening vectors projected from the intrinsic colour of main-sequence stars and giants and are considered to be field stars (main-sequence stars, giants) or 
Class III /Class II sources with small NIR excesses. `T' sources are located redward of 
region `F' but blueward of the reddening line projected from the red end of the T Tauri locus 
of Meyer et al. (1997). These sources are considered to be mostly classical T Tauri stars 
(Class II objects) with large NIR excesses. There may be an overlap in NIR colours of 
Herbig Ae/Be stars and T Tauri stars in the `T' region (Hillenbrand et al. 1992). `P' 
sources are those located in the region redward of region `T' and are most likely Class I 
objects (protostar-like objects; Ojha et al. 2004a). Here it is worthwhile to mention that Robitaille et al. (2006) have recently shown that there is significant overlap between 
protostars and T Tauri stars in colour-colour space.
Comparison of Figs 11a and 11b shows that 
the stars in the cluster region are distributed in a much wider range as compared to stars in the 
reference field. A significant number of the sources in the cluster region exhibit NIR excess 
emission, a characteristics of young stars with circumstellar material. Fig. 11a also shows a distribution of $H\alpha$ emission stars. $H\alpha$ emission stars are distributed in the `T' and `F' regions. 

In Fig. 12a $J,(J-H)$ CMD is plotted for the stars within the cluster region. Fig. 12b shows 
statistically cleaned $J,(J-H)$ CMD obtained using the procedure described in \S 4.3. Using 
the relation ${{A_J}\over {A_V}}=0.265$, ${{A_{J-H}}\over{ A_V}}=0.110$ (Cohen et al. 1981) 
and $A_V=3.1\times E(B-V)$, the isochrones for age 4 Myr by Bertelli et al. (1994) and PMS 
isochrones for ages 1, 10 Myr by Siess et al. (2000) have been plotted assuming a distance 
of 1.0 kpc and extinction $E(B-V)_{min}=1.4$ mag as obtained from the optical data. Probable 
T-Tauri stars, obtained from the $(J-H)/(H-K)$ colour-colour diagram and $H\alpha$ emission 
stars, are shown by open triangles and star symbols respectively and indicate ages $\le$ 1 Myr. The star numbers 3 (ID no. is taken from WEBDA, Sp. type O9.5 V, p=0.98) and 6 (Sp. type B, p=0.99) lie leftward of the MS in the $J, (J-H)$ CMD, whereas on the $V_0, (B-V)_0$ CMD these stars lie on/near the MS. 
We presume that this scatter in $J, (J-H)$ CMD may be due to relatively large errors in NIR data. WEBDA star no. 7 (p=0.72) lies toward the right of 4 Myr isochrone in both the $V_0, (B-V)_0$ and $J, (J-H)$ CMDs. The $(U-B)$ colour for this star is not available therefore an average reddening of the region was applied. Recently Sharma et al. (2007a) have carried out $J$ and $H$ photometry in the central region of Be 59. The  WEBDA star no. 7 is located rightward of the 4 Myr isochrone in their $J, (J-H)$ CMD also. If this star is a cluster member, its evolutionary status does not match with the other massive stars in Be 59. 

The mass of the  probable YSO candidates can be estimated by comparing their locations on the NIR CMD with the evolutionary models of PMS stars. To estimate the stellar masses, the $J$ luminosity is recommended rather than that of $H$ or $K$, as the $J$ band is less affected by the emission from circumstellar material (Bertout et al. 1988). Fig. 13 shows $J,(J-H)$ CMD for probable YSO candidates identified in Fig 11. The solid curve, taken from Siess et al. (2000) denotes the loci of 1 Myr old PMS stars having masses in the range of 0.1 to 3.5 $M_\odot$.  The majority of YSOs have masses in the range 3.0 to 0.1 $M_\odot$.  

\subsection {Initial Mass Function and K-band luminosity function}

The distribution of stellar masses that form in a star-formation event in a given volume of space is called Initial Mass Function (IMF). Young embedded clusters are important  tools to study the IMF as their MF can be considered as the IMF since they are too young to lose a significant number of members either by dynamical or stellar evolution.  The MF is often expressed by a power law,
 $N (\log m) \propto m^{\Gamma}$ and  the slope of the MF is given as:

    $$ \Gamma = d \log N (\log m)/d \log m  $$

\noindent

 where $N  (\log m)$ is the number of stars per unit logarithmic mass interval. The classical value derived by Salpeter  (1955) is $\Gamma = -1.35$. 

With the help of statistically cleaned CMD, shown in Fig. 9,  we can derive the 
MF using the theoretical evolutionary model of Bertelli et al. (1994).
Here we would like to remind that necessary corrections for the incompleteness of the data sample
have been taken into account to get the statistically cleaned CMD. Since 
post-main-sequence age of the cluster is $\sim$ 2 Myr, the stars having $V<15$ mag 
($V_0 < 10; M> 4 M_\odot$) have been considered on the main sequence. For the MS 
stars the LF was converted to MF using the theoretical model by Bertelli et al. (1994) (cf. Pandey et al. 2001, 2005). The MF for PMS stars was obtained by counting the number of stars in various mass bins (shown as evolutionary tracks) in Fig. 9. The MF of the cluster is plotted in Fig. 14. The slope of the mass function $\Gamma$ in the mass range $2.5 <M/M_\odot < 28$ comes out to be $-1.01\pm0.11$, which is  shallower than the Salpeter value (-1.35). A turn-off in the MF can be seen at $\sim$ 2.5 $M_\odot$. In the mass range  $1.5 <M/M_\odot < 2.5$ the mass function is almost flat ($\Gamma \sim$ 0). The shallower
mass function for Be 59 obtained in the present study does not support a universal IMF. 

The K-band luminosity function (KLF) is a powerful tool to investigate the IMF of young embedded stars clusters, therefore during the last decade several studies have focused on determination of the KLF of young open clusters (e.g. Lada \& Lada 2003, Ojha et al. 2004b, Sanchawala et al. 2007). In order to obtain the KLF, we have to examine the effects of incompleteness and field star contamination in our data. The completeness of the data is estimated using the ADDSTAR routine of DAOPHOT as described in \S 2.1. To take into account foreground/background field star contamination we used both the Besan\c con Galactic model of stellar population synthesis (Robin et al. 2003) and nearby reference star field. Star counts are predicted using the Besan\c con model in the direction of the control field. We checked the validity of the simulated model by comparing the model KLF with that of the control field and found that both the KLFs match rather well. An advantage of using the model is that we can separate the foregrounds ($d<1.0$ kpc) and the background ($d>1.0$ kpc) field stars. The foreground extinction is found to be $A_V \sim4.3$ mag (cf. \S 4.2). Model simulations with $A_V$ = 4.3 mag and $d<1.0$ kpc gives the foreground contamination. The background population ($d>1.0$ kpc) was simulated with $A_V$ = 6.7 mag. We thus determined the fraction of the contaminating stars (foreground+background) over the total model counts. This fraction was used to scale the nearby reference field and subsequently the star counts of the modified control field were subtracted from the KLF of the cluster to obtain the final corrected KLF. The KLF is expressed by the following power-law:

%\midskip

${{ \rm {d} N(K) } \over {\rm{d} K }} \propto 10^{\alpha K}$

%\midskip

where ${ \rm {d} N(K) } \over {\rm{d} K }$ is the number of stars per 0.5 magnitude
bin and $\alpha$ is the slope of the power law. Fig. 15 shows the KLF for the cluster region which indicates a slope of $\alpha = 0.27\pm0.02$ which is smaller than the average value of slopes ($\alpha \sim 0.4$) for young clusters (Lada et al. 1991; Lada \& Lada 1995; Lada \& Lada 2003). Smaller values ($\sim 0.3 - 0.2$) have been reported for various young embedded clusters (e.g. Megeath et al. 1996, Chen et al. 1997, Brandl et al. 1999, Ojha et al. 2004b, Leistra et al. 2005, Sanchawala et al. 2007). Since the IMF for Be 59 is found to be shallower ($\Gamma = -1.01 \pm 0.11$) than the Salpeter value, the low value of the slope of the KLF of Be 59 may be an intrinsic property of the Be 59 region. 

Recently we have studied the IMF and KLF for another young cluster (age $\sim$ 4 Myr) NGC 1893 (Sharma et al. 2007b) and it is worthwhile to compare the IMF/ KLF estimates of NGC 1893 with those obtained for Be 59. The mass-luminosity relation  $L_K  \propto m^{\beta}$, KLF and MF i.e. $N (\log m) \propto m^{\Gamma}$ yield relation  ${\alpha}$ = -${\Gamma} \over {2.5{\beta}} $, where $\alpha$, $\Gamma$ and $\beta$ are the slopes of the KLF, IMF and mass-luminosity relation respectively (Lada et al. 1993). The values of $\Gamma$ ($-1.01\pm0.11$ in the mass range $2.5\le M/M_\odot\le28$ for Be 59; $-1.27\pm 0.08$ in the mass range  $0.6\le M/M_\odot\le18$ for NGC 1893) and $\alpha$ ($0.27\pm0.02$ for Be 59; $0.34\pm0.07$ for NGC 1893) yield $\beta \sim 1.5$ for both the clusters which is in between the values reported for massive OB type stars ($\beta = 2$) and PMS stars ($\beta = 1$).

\section {Star formation scenario around the star cluster}

The distribution of YSOs and morphological details of the environment around the cluster containing OB stars can be used to probe the star formation history of the region. 
The OB type stars in the cluster region produce ultraviolet (UV) radiation that ionizes its surroundings, consequently ionization front (IF) is generated. The IF drives a shock into the pre-existing molecular clumps and compresses it, consequently the clumps become gravitationally unstable and collapse to form  next generation stars (Elmegreen 1989, 1998). Fig. 16 shows the spatial distribution of NIR-excess sources and $H\alpha$ emission stars, both being probable T-Tauri stars, on a $50^\prime \times 50^\prime$ DSS-II $R$ band image. The majority of  NIR-excess and $H\alpha$ emission stars are distributed along the elongated morphology of the cluster as well as towards east of the cluster. Yang \& Fukui (1992) studied the S171 region and identified two dense molecular clumps (C1 and C2) near the cluster Be 59. The radial velocities of the clumps V$_{LSR}$= -14.4 km/s (C1) and -16.0 km/s (C2) are comparable with the radial velocity of member stars of Be 59 (V$_{LSR}$= -15.7 km/s), which indicate that the clumps are associated with the cluster. They further found high degree of spatial correlation between the ionized and molecular gas suggesting that these are physically associated. On the basis of the physical association of the molecular clumps, the cluster and the ionized gas, Yang \& Fukui (1992) concluded that a new generation of stars could be expected in the compressed gas layer. 

The clump C1 harbours the bright-rimmed cloud (BRC) BRC1 (cf. Sugitani et al. 1991). BRCs in H{II} regions are potential sites of ongoing star formation due to the radiation driven implosion (RDI), where shock front associated with the ionization front compresses the cloud and causes the BRC to collapse. Morgan et al. (2004, 2006) studied the internal molecular pressure and external ionized boundary layer pressure for 9 BRCs and in the case of BRC1 found that the external pressure is greater than the internal pressure of the molecular cloud which indicates that the ionization front is compressing the molecular gas. In fact, Ogura et al. (2002) have detected 6  H$\alpha$ stars in the vicinity of BRC1 which can be interpreted as second generation of formation of low-mass stars due to RDI.

In Fig. 16 the NVSS (1.4 GHz) radio continuum emission contours (thick line) and 
MSX A-band (8 $\mu$m) MIR contours (thin line) are overlaid on a DSS-II $R$ band image. 
The locations of the IRAS point sources are shown by crosses and the C1 and C2 clumps of 
Yang \& Fukui (1992) are also marked. The peaks of the radio continuum and that of 
the MSX A-band emission around clump C1 coincide very well. The clump C1 also harbours an 
IRAS source which is located in BRC1. Another peak of radio continuum at $ \rm{RA}(2000) = 00^{\rm{h}} 01^{\rm{m}} 12^{\rm{s}}$, $\rm{Dec}.~(2000) = 67^{\circ} 25^{'} 00^{''}$ lies near the clump C2. The emission from MSX A-band is more extended towards the west side of the cluster, whereas it is completely absent towards east of the cluster.

The contribution of UIBs due to polycyclic aromatic hydrocarbons (PAHs) to the mid-infrared emissions in the MSX bands has been studied using the scheme developed by Ghosh \& Ojha (2002). The emission from each pixel is assumed to be a combination of two components, namely the thermal continuum from dust grains (gray body) and the emission from the UIB features in the MSX bands. The scheme assumes a dust emissivity of the power law form $\epsilon_\lambda \propto \lambda^{-1} $ and the total radiance due to the UIBs in band C is proportional to that in band A. A self consistent non-linear chi-square minimization technique is used to estimate the total emission from the UIBs and the warm dust optical depth. The spatial distribution of the UIB emission and optical depth ($\tau_{10}$ at 10 $\mu$m) contour maps  with an angular resolution of $\sim 18^{\prime\prime}$ (for the MSX survey) extracted for the cluster region is shown in Fig. 17. The peaks of the UIB emissions are located at the western border of the radio continuum emission around the clump C1 (see Fig. 17) which indicates an interface between the ionized and molecular gas. The peak of the $\tau_{10}$ optical depth lies at $\sim$ 4 arcmin away towards SW of the cluster center.

Fig. 18 shows the IRAS-HIRES intensity maps for the cluster region where emission shows a strong peak at the clump C1. Another peak towards south-west of the cluster Be 59 can also be noticed. IRAS-HIRES maps were also used to generate the spatial distribution of dust optical depth at 25 $\mu$m  ($\tau_{25}$) and 100 $\mu$m  ($\tau_{100}$), using the procedure given by Ghosh et al. (1993). An emissivity law of $\epsilon_\lambda \propto \lambda^{-1} $ was assumed to generate these maps. The optical depth maps representing warmer (25 $\mu$m) and cooler dust (100 $\mu$m) components are presented in Fig. 19. The warm dust ($\tau_{25}$) is mainly distributed towards east of the clump C1 around the Be 59 cluster position as well as SE and SW of Be 59, whereas the cold dust ($\tau_{100}$) emission is seen west of Be 59 cluster and shows peaks around the clumps C1 and C2. The spatial correlation between the ionized gas, molecular gas and dust around the clumps C1 and C2 suggest that these clumps may be the locations of second generation star formation as suggested by Yang \& Fukui (1992). The morphology of the MSX A-band as well as IRAS intensity maps implies that the HII region is ionization bounded towards the western side of the cluster. 

A comparison is also made between the $\tau_{10}$ maps generated from the higher angular resolution MSX maps (Fig. 17) and that based on the IRAS HIRES maps at 12 and 25 $\mu$m. The later ($\tau_{25}$) is shown in Fig. 19 which was scaled to 10 $\mu$m assuming a $\lambda^{-1}$ emissivity law to compare with the MSX $\tau_{10}$ map. The peak optical depth is 5.49 $\times 10^{-5}$ for the map based on MSX. The corresponding value from the IRAS-HIRES maps is 2.73 $\times 10^{-5}$. These derived values are in reasonable agreement considering that they are based on instruments with very different angular resolutions. The difference in the peak values of $\tau_{10}$ may be a result of the following effects: beam dilution, clumpy interstellar medium and the contribution of UIB emission. In particular the contribution of the UIBs emission has been removed in generating the optical depth in the modeling of the MIR emission from the MSX bands, however the IRAS-HIRES maps represent the emission from dust and UIB carriers (e.g. the emission from the UIB features falls within the IRAS 12 $\mu$m band ($\sim$ 8-15 $\mu$m)). The overestimate of thermal emission from the grains at 12 $\mu$m band due to UIB features would imply a higher effective colour temperature T(12/25) for the grains. This would then affect the inferred optical depth (by reduction in its value).

The $(J-H)/(H-K)$ colour-colour diagrams for the probable YSOs lying around the clumps C1 and C2 are shown in Fig. 20. Since the clump regions are associated with higher extinction, we also used data of those stars having photometric errors in the range 0.1-0.2 mag to improve the statistics. The stars having error larger than 0.1 mag are shown by filled triangles in Fig. 20. A comparison of Fig. 20 with the cluster region colour-colour diagram (Fig. 11) indicates that stars lying in the clump C2 have higher extinction and relatively larger NIR excess suggesting that these stars may be younger than the stars in the  clump C1 and cluster region. If this is true, then star formation in the clump regions might have taken place as a result of triggering as suggested by Yang and Fukui (1992).

To study the age sequence, if any, we selected four sub-regions (regions 1-4) in/around the cluster region and a field region (region 5) outside the cluster as shown in Fig. 21. Fig. 22 shows $(J-H)/(H-K)$ colour-colour diagrams of YSOs in the four sub-regions  and  the field region. The NIR excess stars and H$\alpha$ stars are shown by open triangles and star symbols respectively. The statistics is given in Table 5 which manifests that regions 1 and 4 have higher percentage of YSOs as compared to the regions 2 and 3 which lie in the cluster.  This indicates that the regions 1 and 4 
should be younger than the cluster. Fig. 23 shows dereddened $V_0,(V-I)_0$ CMDs for the probable YSOs (NIR excess and $H\alpha$ emission stars) along with stars on/near the MS (filled circles) in regions 1, 2 and 3. The $(J-H)/ (H-K)$ TCD was used to estimate $A_V$ for each YSO by tracing back to the intrinsic line by Meyer et al. (1997) along the reddening vector (for details see Ogura et al. 2007). Probable MS stars are dereddened individually as described in $\S$4.3. The YSOs in the cluster Be 59 (regions 2 and 3) have ages $<$1 Myr to 2 Myr, whereas all the YSOs except one in region 1 have ages $<$1 Myr. Because of high extinction in region 4 we could not get a sufficient number of $V$ and $I$ observations of $H\alpha$ and NIR excess stars. The majority of probable PMS stars (from statistically cleaned CMD, see \S 4.3) in the cluster region have ages between 1-5 Myr. The average age of the massive stars in the cluster region is estimated to be $\sim$2 Myr. Presence of significant number (7/12) of YSOs in the central region (region 2) of the cluster with ages $<$1 Myr indicates a multi-epoch star formation in the cluster. 

Fig. 24 shows $J,(J-H)$ CMDs of the stars in four sub-regions  having NIR excess and $H\alpha$ emission. The isochrones for 4 Myr by Bertelli et al. (1994) and PMS isochrones for 1 and 10 Myr by Siess et al. (2000) are also plotted for $E(B-V)=1.4$ mag and distance of 1.0 kpc. Since stars in the field region have $(J-H)<1.05$ (cf. Fig. 11b), the stars in the sub-regions having $(J-H)>1.05$ are considered as probable members of the Be 59 star forming region. The CMDs show that region 4 has more NIR excess stars in comparison to the cluster (regions 2 and 3) supporting the youth of the region. 
The mass of the  probable YSO candidates can be estimated by comparing their locations on the CMD with the evolutionary models of PMS stars. The continuous oblique reddening lines denote the positions of PMS stars (age $\sim $ 1 Myr) of 3.5 and 0.8 $M_\odot$ respectively. The YSOs in all the four sub-regions have masses in the range $\sim 3.0 - 0.8 M_\odot$. A comparison of the distribution of stars in Fig. 24 indicates that the region 2 (central region of Be 59) has relatively more massive stars (of 24 stars, 6 have masses $> 2.0 M_\odot$) in comparison to region 4 (1 star out of 15 stars has mass $> 2.0 M_\odot$).

Assuming the number density of  the molecular gas and temperature of ionized gas as $3.5\times 10^2$ cm$^{-3}$ and $\sim 10^4$ K respectively, Yang \& Fukui (1992) have estimated the shock velocity as $\sim$ 3 km/s. Since, the projected distance of clumps C1 and C2 from the cluster center is $\sim$ 3 pc, the triggering shock would take  $\sim$ 1 Myr to reach the clumps.  This indicates that the age difference between the PMS stars towards the clumps and triggering stars in the cluster center should be $\sim$ 1-2 Myr. The age of young clusters are usually derived by fitting the post-main sequence evolutionary tracks to the earliest members if significant evolution has occurred or by fitting the theoretical PMS isochrones  to the low-mass contracting population. The $(J-H)/(H-K)$ CC diagrams (Fig. 22), CMDs (Fig. 24) and distribution of YSOs towards the clumps (Table 5)  indicate that the YSOs around the clumps are young (age $\lesssim$ 1 Myr) as compared to those in the cluster. The probable PMS stars in the cluster region show an age spread of $\lesssim$ 1 - 5 Myr (Fig. 9), whereas NIR excess stars and H$\alpha$ emission stars have age $\lesssim$ 1 - 2 Myr. The average post-main sequence ages of massive stars are estimated as $\lesssim$ 2 Myr. Here it is worthwhile to mention that massive stars are not significantly evolved, therefore age determination of post-age main sequence stars may be less accurate in comparison to the PMS stars. On the basis of above discussions, the expected age difference between the cluster population and YSOs towards clumps is estimated to be $\sim$ 1-2 Myr. This seems to support the triggered star formation activity towards the clumps due to massive stars in the cluster region.

\section {The NIR excess and H$\alpha$ emission stars: statistics}

To study the effect of environment on the disk around the YSOs we present statistics of NIR excess stars and H$\alpha$ emission stars of the clusters Be 59 and NGC 1893 in Table 6. The data for NGC 1893 have been taken from Sharma et al. (2007b). 
The number of PMS stars in the range $M_V \sim 0 - 5$ mag has been estimated from the statistically cleaned CMD as discussed in \S  4.3. 
The statistics indicates that the inner region of the cluster NGC 1893 shows a less percentage of PMS-NIR excess stars as compared to the outer region, whereas both inner and outer regions of Be 59 show rather a similar distribution of PMS-NIR excess stars.
In the case of Be 59 the percentage of PMS-NIR excess stars in the inner region is higher than in the case of NGC 1893. 

Strong H$\alpha$ emission in TTSs generally has been attributed to accretion process, whereas weak emission is believed to arise from chromospheric activity. The TTSs in the young cluster IC 5146 show that NIR excess and 
equivalent width  EW(H$\alpha$) are well correlated (Herbig \& Dahm 2002, Dahm 2005). In Fig. 25 we plot EW(H$\alpha$) against NIR excess $\Delta (H-K)$, defined as the horizontal displacement from the reddening vector at the boundary of `F' and `T' regions (see Fig. 11). Apparently there does not seem any trend as statistics is poor, however lower profile of the distribution seems to follow the dashed line which is an eye estimated lower profile for the data points shown in figure 10 of Herbig \& Dahm (2002). Here we have applied a correction of -0.25 to the $\Delta (H-K)$ values of  Herbig \& Dahm (2002) as they measured $\Delta (H-K)$ from the leftward reddening vector of `F' region. 

Fig. 26 shows a plot of EW(H$\alpha$) as a function of $J$ magnitude, which indicates that faint H$\alpha$ stars ($J>14$, mass $\lesssim 0.8 M_\odot$) have higher EW(H$\alpha$) than the brighter stars (mass $\sim 1 -3$ $M_\odot$). This effect partly may be due to magnitude dependence of EW(H$\alpha$), however it appears accretion activity among low-mass members of Be 59 has not subsided. About 32\% and 62\%  of H$\alpha$ stars of Be 59 and NGC 1893, respectively, show NIR excess in $(H-K)$ colour (see Table 7). For comparison, this ratio  is 66\% in IC 5146 (TTSs age $<$ 0.7 Myr), 27\% in NGC 2264 (TTSs age $\sim$ 1.1 Myr) and 4\% in NGC 2362 (TTSs age $\sim$ 1.4 Myr) (Dahm 2005). The presence of H$\alpha$ emission and NIR excess in  TTS population indicates that the inner disk regions of low mass stars are still not dissipated.

Table 8 gives the statistics of the YSOs as a function of age. We selected a sample of probable PMS stars in the range $M_V \sim 0 -5$ mag from the statistically cleaned CMDs (cf \S 4.3). The percentage of NIR excess and  H$\alpha$ emission stars decreases with age. In the cluster Be 59 $\sim 29\%$ and $71\%$ of the PMS stars (having age $\le$ 1 Myr) show NIR excess and H$\alpha$ excess emission respectively, whereas in the case of NGC 1893 this fraction is smaller ($10\%$ and $12\%$ respectively). 
 
Finally we compare the surface densities of H$\alpha$ emitters in various young star clusters. The slitless spectroscopy of Be 59 and NGC 1893 (Sharma et al. 2007b) is expected to be complete upto $M_V$=5. The surface densities of  H$\alpha$ emitters in clusters Be 59 (area $\simeq$ 26.6 pc$^2$) and NGC 1893 (area $\simeq$ 101 pc$^2$) are found to be 1.0 star/pc$^2$ and 0.2 star/pc$^2$ respectively. The surface densities of  H$\alpha$ emitters in young  clusters NGC 2264 (area $\simeq$ 52 pc$^2$) and NGC 2362 (area $\simeq$ 22.4 pc$^2$), estimated for magnitude limit $M_V$=5 from Dahm (2005) are 1.4 star/pc$^2$ and 0.2 star/pc$^2$ respectively. The density of H$\alpha$ emission stars in central region of both the clusters Be 59 and NGC 1893 is higher as compared to the outer region of the clusters (see Table 7).

\section {Summary}

The star formation scenario around young star cluster Be 59 is studied using the  optical $UBVI_C$  
photometric data, slitless spectroscopy along with archival data from the surveys such as 2MASS, MSX, IRAS and NVSS.  The post-main-sequence age and distance of the cluster are found to be $\sim 2$ Myr and  1.00 $\pm 0.05$ kpc respectively. YSOs in the Be 59 region have been identified using NIR two colour diagram and  $H\alpha$ emission stars. Following are the main results:

1) The interstellar extinction in the cluster region is found to be variable with $E(B-V)_{min} \simeq 1.4$ mag, $E(B-V)_{max} \simeq 1.8$ mag. The ratio of total-to-selective extinction in the cluster region, $R_{cluster}$ is estimated as $3.7 \pm 0.3$ which indicates an anomalous reddening law consequently a larger grain size in the Be 59 region.

2) The cluster Be 59 contains OB stars having mean age $\sim$ 2 Myr along with probable PMS stars having age 1-5 Myr. 
A slitless spectroscopic survey of Be 59 region identifies 48 H$\alpha$ emission stars. 
The ages of the YSOs (H$\alpha$ emission stars and NIR excess stars) are found to be in the range of $<1$ Myr to 2 Myr.

3) The slope of the initial mass function $`\Gamma$' in the mass range $2.5 < M/M_\odot \le 28$ is found to be $-1.01\pm0.11$ which is shallower than the Salpeter value (-1.35), whereas in the mass range $1.5 < M/M_\odot \le 2.5$ the slope is almost flat. The slope of the K-band luminosity function is estimated as $0.27\pm0.02$ which is smaller than the average value ($\sim$0.4) reported for young clusters.

4) Approximately $32\%$ of H$\alpha$ emission stars of Be 59 exihibit NIR excess indicating that their 
inner disk has not yet dissipated.  
The inner region of the cluster NGC 1893 has smaller percentage of PMS-NIR excess stars as compared to outer region. In the case of Be 59 the percentage of PMS-NIR excess stars in the inner as well as in the outer region is same. 

5) The $(J-H)/(H-K)$ colour-colour diagrams and $J,(J-H)$ CMD for the stars lying towards the molecular 
clumps C1 and C2 (cf. Yang and Fukui, 1992) indicate that these stars have higher extinction and relatively larger NIR excess suggesting that they may be younger than the stars in the cluster region. If this is true then star formation towards the clump regions might be triggered by OB stars in the cluster region.

6) The surface densities of H$\alpha$ stars, having $M_V \le$5, for clusters Be 59 and NGC 1893 are found to be 1.0 star/pc$^2$ and 0.2 star/pc$^2$ which are comparable to the surface densities (for the same magnitude limit) for young clusters NGC 2264 (1.4 star/pc$^2$) and NGC 2362 (0.2 star/pc$^2$) estimated from the work Dahm (2005). 

\section*{Acknowledgments}
Authors are thankful to the anonymous referee for useful comments.
The observations reported in this paper were obtained using the Kiso Schmidt (Japan) and  2 meter HCT at IAO, Hanle, the high altitude station of Indian Institute of Astrophysics, Bangalore. We are also thankful to the Kiso observatory and IAO for allotting observing time. We thank the staff at Kiso (Japan) and IAO, Hanle/its remote control station at CREST, Hosakote for their assistance during the observations. This publication makes use of data from the Two Micron All Sky Survey, which is a joint project of the University of Massachusetts and the Infrared Processing and Analysis Center/California Institute of Technology, funded by the National Aeronautics and Space Administration and the National Science Foundation. AKP is thankful to the National Central University, Taiwan and TIFR, Mumbai for the financial support during his visit to NCU and TIFR respectively. AKP and KO acknowledge the support given by DST (India) and JSPS (Japan) to carry out the wide field CCD photometry around open clusters. 
We thank Annie Robin for letting us use her model of stellar population synthesis.

\begin{table*}
\centering
\begin{minipage}{140mm}
\caption{Log of observation.}
\begin{tabular}{@{}lrr@{}}
\hline
\hline
Telescope/Date & Filter& Exp. (sec) $\times$ No. of frames\\
\hline
Kiso Schmidt, Japan&&\\
19 November 2001&$U$   &  $180\times15$\\
~~~~~~~~~~"&$B$   &  $180\times10,60\times16,10\times6$\\
~~~~~~~~~~"&$V$   &  $60\times16,10\times9$\\
20 November 2001 &$I_C$   &  $60\times3,10\times10$\\\\
Himalayan Chandra Telescope, IIA&&\\
22 September 2004, and 7,8,9 January 2005 &H$\alpha$ with grism in & ($420\times3$)$\times 9~ positions$ \\
~~~~~~~~~~"&H$\alpha$ without grism & ($60\times3$)$\times 9~ positions$\\

\hline
\end{tabular}
\end{minipage}
\end{table*}

\begin{table*}
\centering
\begin{minipage}{140mm}
\caption{ $UBVI_c$ photometric data of the stars in the field of Be 59. $X$, $Y$ are
the pixel coordinates. Radius is with respect to the center $X$ = 1128, 
$Y$ = 1024 ($\rm{\alpha}_{2000} = 00^{\rm{h}} 02^{\rm{m}} 10^{\rm{s}}.4$, 
$\rm{\delta}_{2000} = 67^{\circ} 25^{'} 10^{''}$). One pixel corresponds to 1.5 arcsec. The complete
table is available in electronic form only.}
\begin{tabular}{@{}rrrrrrrrrrr@{}}
\hline
\hline

 Star No. &   $X$  &   $Y$   &   $V$   &  $B-V$  &  $V-I$  &  $U-B$  & Radius   & $\alpha_{2000}$    & $\delta_{2000}$ \\
          & (pixel)&(pixel)  &(mag)    &(mag)    & (mag)    & (mag)    & (pixel) & (h:m:s)   &  (d:m:s)\\
\hline
1        &1128.92&  1023.27& 19.126& ------&  ------& ------&     1.17&   00:02:10.13&   67:25:08.6\\
2        &1117.29&  1028.99& 18.508& ------&   3.039& ------&    11.81&   00:02:13.15&   67:25:17.2\\
3        &1115.29&  1020.07& 11.301&  1.143&   1.426&  0.008&    13.30&   00:02:13.69&   67:25:03.9\\
4        &1134.18&  1011.54& 17.654&  2.193&   3.024& ------&    13.90&   00:02:08.78&   67:24:50.9\\
5        &1143.74&  1012.27& 19.475& ------&   3.622& ------&    19.63&   00:02:06.29&   67:24:51.9\\
6        &1147.39&  1014.66& 18.303&  2.421&   3.159& ------&    21.52&   00:02:05.33&   67:24:55.5\\
7        &1137.94&  1045.48& 13.873&  1.241&   1.648&  1.181&    23.66&   00:02:07.74&   67:25:41.8\\
8        &1146.54&  1040.00& 18.657&  1.723&   2.554& ------&    24.48&   00:02:05.50&   67:25:33.5\\
\hline
\end{tabular}
\end{minipage}
\end{table*}

\begin{table*}
\centering
\caption{Detected $H\alpha$ emission stars.}
\begin{tabular}{rrrrrrrr}
\hline
\hline

ID  & $\alpha_{2000}$   & $\delta_{2000}$  & EW         & $V$    & $B-V$     &   $V-I$  \\
        &(h:m:s)        &  (d:m:s)         &($\rm \AA$) &(mag)   & (mag)     & (mag)    \\
\hline

 1   &00:00:03.93 & +67:35:55.1 &15  &  19.218 &  2.121 &  2.794\\
 2   &00:00:08.23 & +67:35:47.1 &12  &    --   &    --  &   --  \\
 3   &00:00:12.02 & +67:34:42.1 &76  &    --   &    --  &   --  \\
 4   &00:00:18.68 & +67:28:09.8 &19  &  19.308 &    --  &  3.023\\
 5   &00:00:20.40 & +67:27:17.0 &47  &    --   &    --  &   --  \\
 6   &00:00:53.93 & +67:29:25.0 &3   &    --   &    --  &   --  \\
 7   &00:01:38.76 & +67:28:00.5 &--  &    --   &    --  &   --  \\
 8   &00:01:45.77 & +67:17:41.3 &55  &  19.084 &    -   &  3.054\\
 9   &00:01:48.40 & +67:27:28.5 &25  &  18.738 &  2.407 &  3.235\\
10   &00:01:50.02 & +67:30:03.7 &--  &    --   &    --  &   --  \\
11   &00:01:54.49 & +67:30:17.6 &56  &    --   &    --  &   --  \\
12   &00:01:54.99 & +67:27:12.6 &21  &    --   &    --  &   --  \\
13   &00:01:57.89 & +67:23:18.5 &13  &  17.552 &  2.372 &  2.840\\
14   &00:01:58.09 & +67:22:38.0 &--  &    --   &    --  &   --  \\
15   &00:01:59.27 & +67:24:34.6 &16  &    --   &    --  &   --  \\
16   &00:02:00.39 & +67:23:57.4 &8   &  15.900 &  1.837 &  2.236\\
17   &00:02:02.05 & +67:23:05.5 &11  &  18.667 &  2.367 &  3.349\\
18   &00:02:02.29 & +67:22:36.1 &--  &    --   &    --  &   --  \\
19   &00:02:02.55 & +67:22:38.6 &--  &    --   &    --  &   --  \\
20   &00:02:02.56 & +67:36:46.8 &--  &    --   &    --  &   --  \\
21   &00:02:04.68 & +67:36:49.1 &--  &    --   &    --  &   --  \\
22   &00:02:06.31 & +67:24:51.9 &--  &  19.475 &    --  &  3.622\\
23   &00:02:07.82 & +67:28:38.0 &11  &  19.528 &    --  &  3.333\\
24   &00:02:15.08 & +67:29:27.0 &25  &  18.268 &  2.417 &  2.754\\
25   &00:02:16.33 & +67:23:37.0 &19  &  18.991 &    --  &  3.476\\
26   &00:02:19.64 & +67:18:13.3 &8   &  18.585 &  3.117 &  3.467\\
27   &00:02:24.51 & +67:28:20.2 &8   &  17.519 &  2.530 &  3.132\\
28   &00:02:37.34 & +67:32:04.4 &15  &    --   &    --  &   --  \\
29   &00:02:43.20 & +67:34:56.7 &14  &  19.249 &    --  &  3.212\\
30   &00:02:46.47 & +67:19:09.0 &54  &    --   &    --  &   --  \\
31   &00:02:48.86 & +67:29:01.7 &6   &  18.347 &  2.445 &  3.110\\
32   &00:02:57.61 & +67:34:42.6 &14  &  17.751 &  2.142 &  2.540\\
33   &00:03:10.93 & +67:35:21.7 & 6  &  18.007 &  2.472 &  2.997\\
34   &00;03;14.72 & +67;33;08.4 & -- &    --   &    --  &   --  \\
35   &00:03:18.77 & +67:30:59.5 & -- &    --   &    --  &   --  \\
36   &00:03:19.48 & +67:25:16.9 & 50 &    --   &    --  &   --  \\
37   &00:03:27.33 & +67:37:56.9 & 14 &  18.676 &  2.350 &  3.031\\
38   &00:03:28.36 & +67:25:53.3 & 11 &  17.742 &  2.558 &  3.055\\
39   &00:03:35.14 & +67:38:25.4 & -- &    --   &    --  &   --  \\
40   &00:03:49.93 & +67:23:47.1 & 8  &  18.789 &  2.543 &  3.158\\
41   &00:03:51.49 & +67:31:22.3 & 9  &  19.545 &  1.926 &  2.573\\
42   &00:04:09.21 & +67:22:37.7 & 18 &  17.252 &  2.356 &  2.888\\
43   &00:04:13.68 & +67:24:45.7 & 55 &  18.563 &  2.876 &  3.292\\
44   &00:04:18.14 & +67:24:33.5 & -- &    --   &    --  &   -- \\
45   &00:04:18.76 & +67:26:27.1 & -- &    --   &    --  &   -- \\
46   &00:04:20.31 & +67:28:18.1 & 48 &    --   &    --  &   -- \\
47   &00:04:37.51 & +67:28:55.5 & 41 &    --   &    --  &   -- \\
48   &00:04:09.21 & +67:22:37.7 & 22 &    --   &    --  &   -- \\

\hline
\end{tabular}
\end{table*}

\begin{table*}
\centering
\begin{minipage}{140mm}
\caption{List of IRAS point sources.}
\begin{tabular}{@{}rrrrrrrr@{}}
\hline
\hline

   IRAS PSC & $\alpha_{2000}$&$\delta_{2000}$& $F_{12}$   &    $F_{25}$&      $F_{60}$&     $F_{100}$\\
            &(h:m:s)&(d:m:s)&          (Jy)&          (Jy)&          (Jy)&          (Jy)\\
\hline
 23568+6706&  23:59:25.81 & 67:23:39.1&  5.80&   9.75&   184.0&  841.0\\
 00025+6708&  00:05:07.39 & 67:24:45.0&  0.79&   1.38&   16.6 &  73.9\\
 00002+6647&  00:02:47.50 & 67:03:47.2&  0.77&   1.31&   13.0 &  59.5\\
 00034+6658&  00:06:06.31 & 67:15:27.0&  0.90&   1.72&   18.4 &  68.5\\
\hline

\end{tabular}
\end{minipage}
\end{table*}

\begin{table*}
\centering
\caption{Statistics of YSOs in four sub-regions. Numbers given in parentheses are in 
percentage of probable cluster members. }
\begin{tabular} {r r c c l}
\hline
\hline                                                 
Region    &  Total  &        NIR excess  & $H\alpha$    &    All YSOs     \\
          &  stars  &        stars       & stars      &                 \\
\hline                                                 
  1       &  288    &         15 (50)    &  9 (41)    &      18 (82)$^a$  \\
  2       &  686    &         33 (7)     & 16 (4)     &      40 (10)$^b$  \\
  3       &  358    &          9 (5)     &  6 (7)     &      11 (12)  \\
  4       &  229    &       17 (100)$^*$ & 4 (100)$^*$&     16 (100)$^c$ \\
Field     &  266    &          4         &            &                 \\
\hline
\end{tabular}

*: Total stars in the region 4 are less than the field region therefore\\
all the stars in region 4 are considered as NIR excess/$H\alpha$ stars.\\

a: Two $H\alpha$ stars have NIR excess\\
b: Five $H\alpha$ stars have NIR excess\\
c: One $H\alpha$ star has NIR excess\\

\end{table*}

\begin{table*}
\centering
\caption{Statistics of PMS stars (having $M_V \sim 0 - 5$ mag), NIR excess and  H$\alpha$ stars in clusters Be 59 and NGC 1893}
\begin{tabular} {r r c c l}
\hline
\hline                                                 
Population        &      \multicolumn{2}{c}{ Be 59}    &  \multicolumn{2}{c}{NGC 1893}     \\
                  & $r<5^\prime$ & $5^\prime<r<10^\prime$ & $r<3^\prime$& $3^\prime<r<6^\prime$  \\
\hline                                                 
PMS stars  &  186      &    219      &       99      &     167 \\
&&&& \\
NIR excess stars       & 11      &   15     &   3      &    10  \\
&&&& \\
\% of PMS stars having NIR excess& 5.9     &   6.8      &      3.0     &    6.0 \\
&&&& \\
H$\alpha$ stars       & 6      &   8      &  5      &    4 \\
&&&& \\
\% of PMS stars having H$\alpha$ emission& 3.2      &   3.7      &      5.0    &   2.4 \\
\hline
\end{tabular}
\end{table*}

\begin{table*}
\centering
\caption{Statistics of H$\alpha$ stars showing NIR excess in clusters Be 59 and NGC 1893. This table lists all the H$\alpha$ stars which have $JHK_s$ data.}
\begin{tabular} {r r c c l}
\hline
\hline                                                 
Population        &      \multicolumn{2}{c}{ Be 59}    &  \multicolumn{2}{c}{NGC 1893}     \\
                  & $r<5^\prime$ & $5^\prime<r<10^\prime$ & $r<3^\prime$& $3^\prime<r<6^\prime$  \\
\hline   

$H\alpha$ stars  \\ 
(having $JHK_s$ data)&  12      &    10      &       7      &     9 \\
&&&& \\
$H\alpha$  star having  \\
NIR excess&   4      &     3      &       4      &     6 \\
&&&& \\
\% of $H\alpha$ stars with\\
NIR excess  &   33     &     30     &       57     &     67 \\
&&&& \\
Density of $H\alpha$ stars   \\ 
($stars/pc^2$)  &  2.3     &   0.6      &      0.4     &    0.2 \\
\hline
\end{tabular}
\end{table*}

\begin{table*}
\centering
\caption{Statistics of YSOs (having $M_V \sim 0-5$ mag) in clusters Be 59 and NGC 1893. Numbers given in parentheses are in percentage of PMS stars. }
\begin{tabular} {c c c c c c c c }
\hline
\hline                                                 
Age range &      \multicolumn{3}{c}{ Be 59}                      &&              \multicolumn{3}{c}{NGC 1893}     \\
(Myr)     &  PMS stars  &    NIR excess stars  & $H\alpha$ stars && PMS stars  &    NIR excess stars  & $H\alpha$ stars \\
\hline                                                             
$<$1      &   14        &      4(29)           &  10(71)         &&  50        &      5(10)           &    6(12)   \\
1-2       &   8         &      1(13)           &   2(25)         &&  60        &      3(5)            &    2(3)    \\
2-5       &   22        &      2(9)            &   1(5)          &&  49        &      0(-)            &    1(2)    \\
5-10      &   13        &      1(8)            &   1(8)          &&  7         &      0(-)            &    1(14)   \\
\hline
\end{tabular}
\end{table*}

\clearpage

\begin{figure*}
\includegraphics[height=7cm,width=6cm]{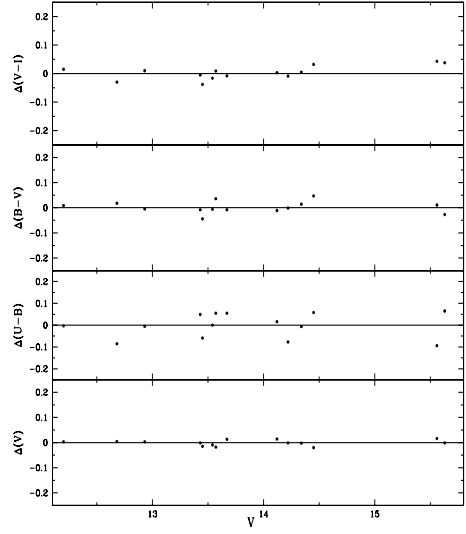}
\caption{Residuals between standard and transformed magnitudes and colour of standard
stars plotted against standard $V$ magnitude.}
\end{figure*}

\begin{figure*}
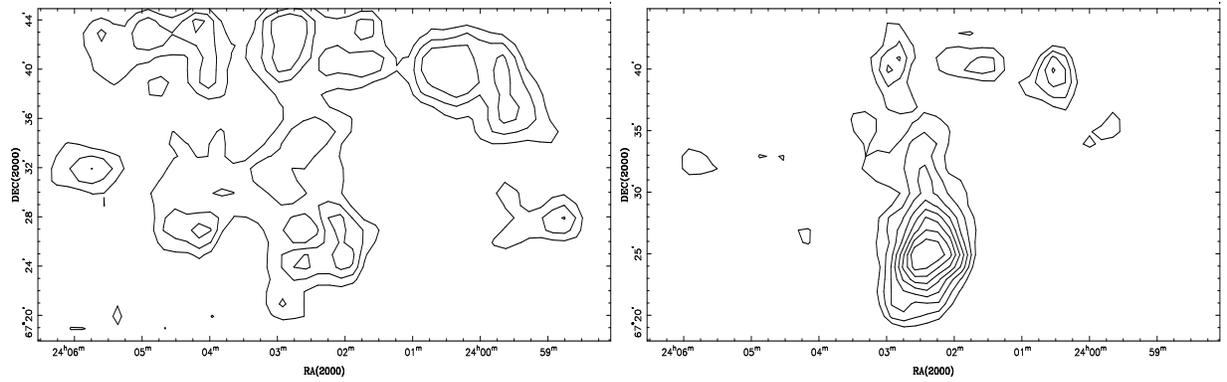

\vbox{
\includegraphics[height=8cm,width=5cm,angle=-90]{fig02a.ps}
\includegraphics[height=8cm,width=5cm,angle=-90]{fig02b.ps}
}
 \caption{({\it left panel}) Isodensity contours for stars having $V\le18$ mag; ({\it right panel}) 
Isodensity contours for 2MASS stars in $K_{\rm s}$ band. The contours are plotted above 3 sigma levels. 
For optical data, the contours have step size of 5 star/pc$^2$ with the lowest contour representing
15 stars/pc$^2$. The maximum stellar density is ~33 stars/pc$^2$.
For 2MASS data, the contours have step size of 5 stars/pc$^2$ with lowest contour representing
27 stars/pc$^2$. The maximum stellar density is ~70 stars/pc$^2$.
The abscissa and the ordinate are for the J2000 epoch. }

\end{figure*}

\begin{figure*}
\includegraphics[height=8cm,width=8cm,angle=-90]{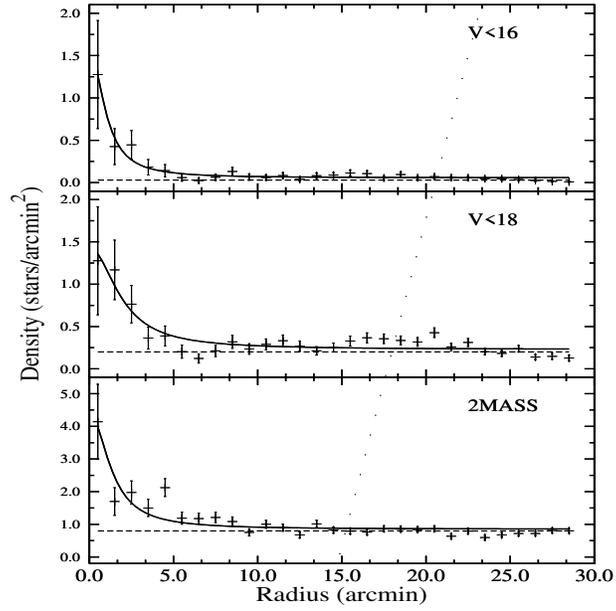}
\caption{Radial density profiles of probable MS stars of the cluster for different optical magnitude levels and 2MASS data. The continuous curve shows a least-square
fit of the King (1962) profile to the observed data points.
The error bars represents $\pm\sqrt{N}$ errors. The dashed line indicates the density
of field stars. }
\end{figure*}

\begin{figure*}
\includegraphics[height=8cm,width=8cm]{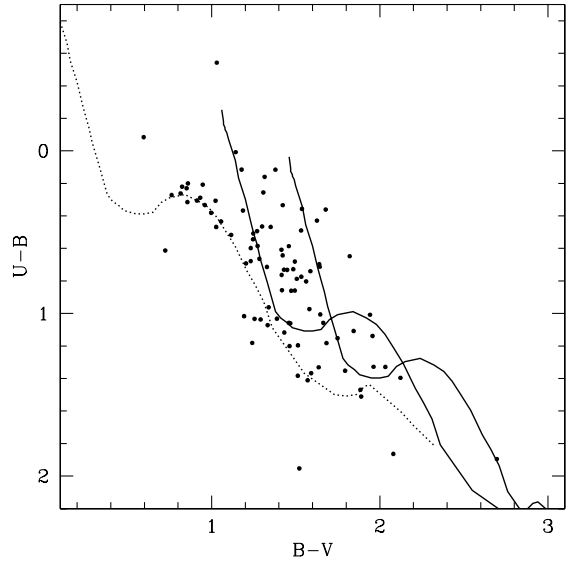}
\caption{The $(U-B)/(B-V)$ colour-colour diagram for the star within the region $r<r_{cl}$. The continuous curves 
represent ZAMS by Schmidt-Kaler (1982) shifted along the reddening slope of 0.72 
for $E(B-V)$ = 1.4 and 1.8 mag respectively. The dashed curve represents ZAMS shifted for 
the foreground reddening $E(B-V)$ = 0.4 mag. }
\end{figure*}

\begin{figure*}
\includegraphics[height=8cm,width=5cm,angle=-90]{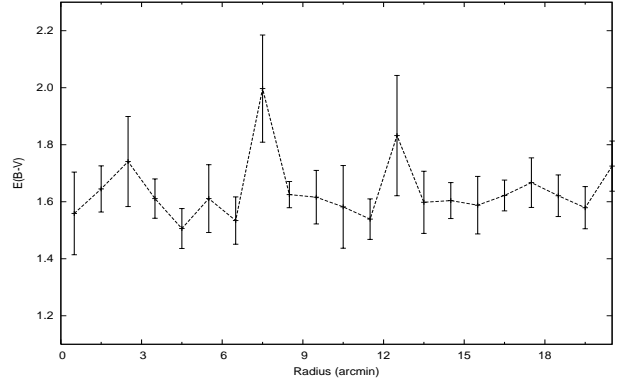}
\caption{Radial distribution of the reddening from center to outer region of the cluster. Error bars represent standard errors.}
\end{figure*}

\begin{figure*}
\includegraphics[height=16cm,width=10cm,angle=-90]{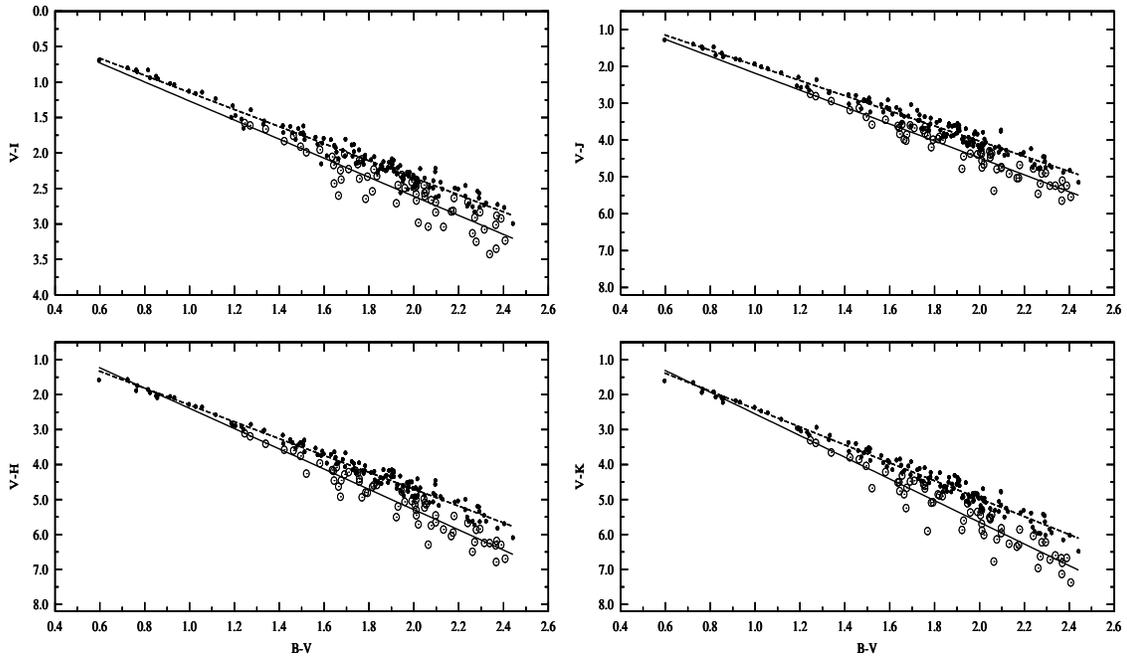}
\caption{$(V-I), (V-J), (V-H), (V-K)$ vs. $(B-V)$ two colour diagrams within cluster region
($r<r_{cl}$). Open and filled circles, respectively, represent probable cluster members with anomalous reddening and field stars with normal reddening. Straight lines show least square fit to the data.}
\end{figure*}

\begin{figure*}
\includegraphics[height=14cm,width=16cm]{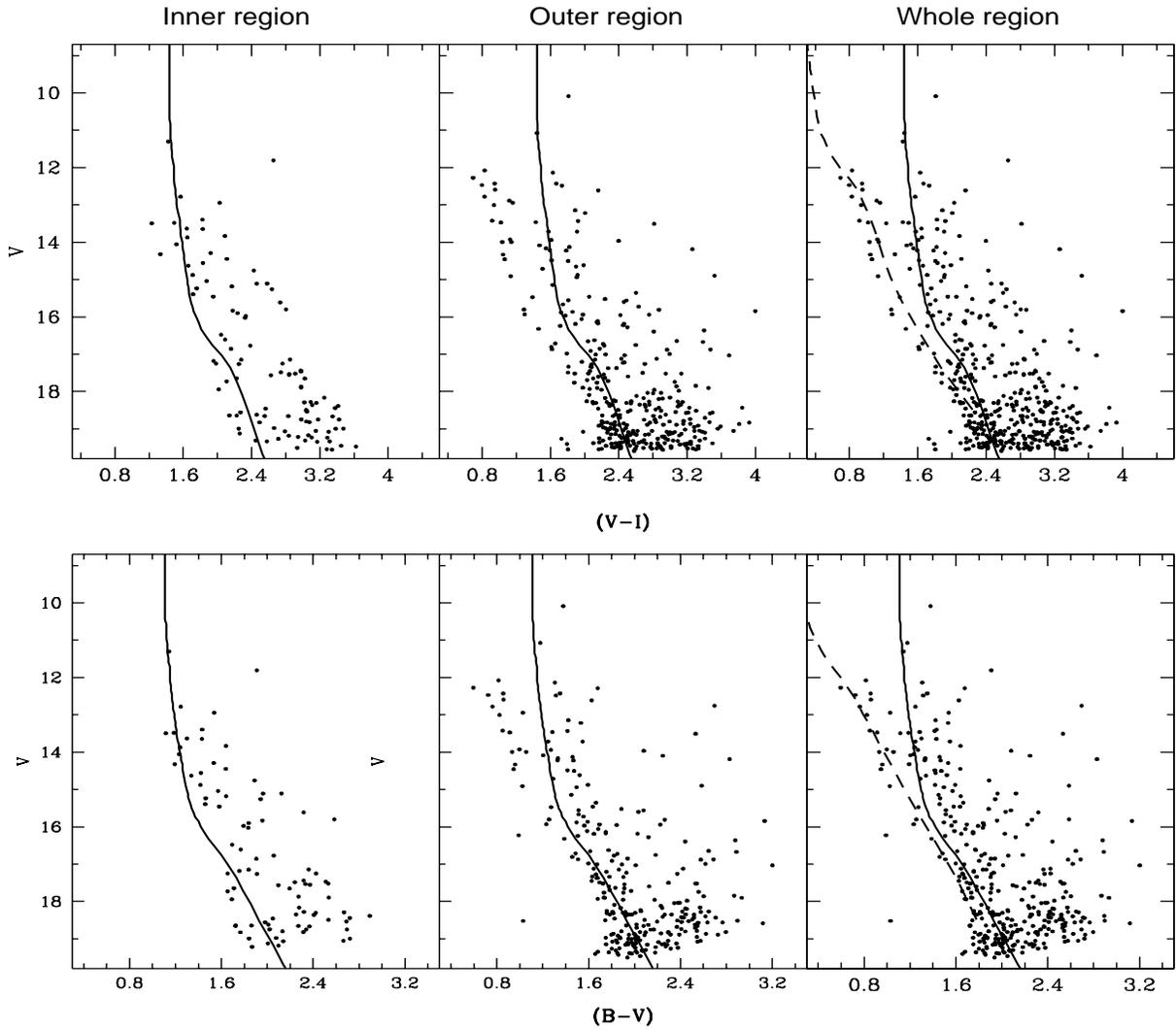}
\caption{ The CMDs for stars lying in the two subregions ($r \le 3^\prime$ and $ 3^\prime < r \le 10^\prime$) and whole cluster region.
The isochrone by Bertelli et al. (1994) for solar metallicity and age of 4 Myr are 
also plotted for distance modulus of 14.35 mag and minimum reddening $E(B-V)=1.4$ mag. The dashed curve is ZAMS for $E(B-V)$=0.4 and distance modulus of 9.60 mag.  }
\end{figure*}

\begin{figure*}
\includegraphics[height=6cm,width=8cm]{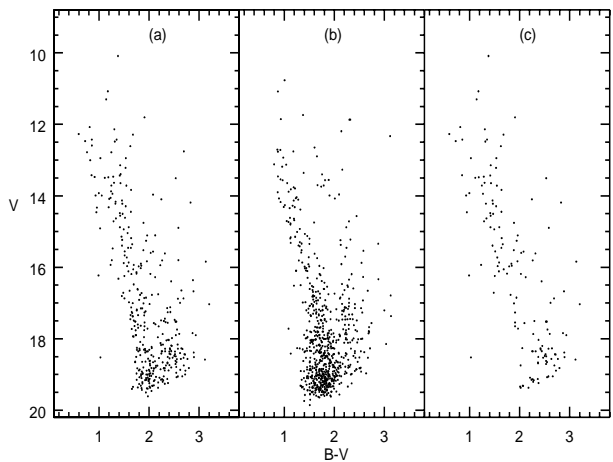}
\caption{$V,(B-V)$ CMDs for (a) stars in the cluster region ($r\le 10^\prime$), (b) stars in the field 
region and (c) statistically cleaned CMD of the cluster region. }
\end{figure*}

\begin{figure*}
\includegraphics[height=8cm,width=8cm]{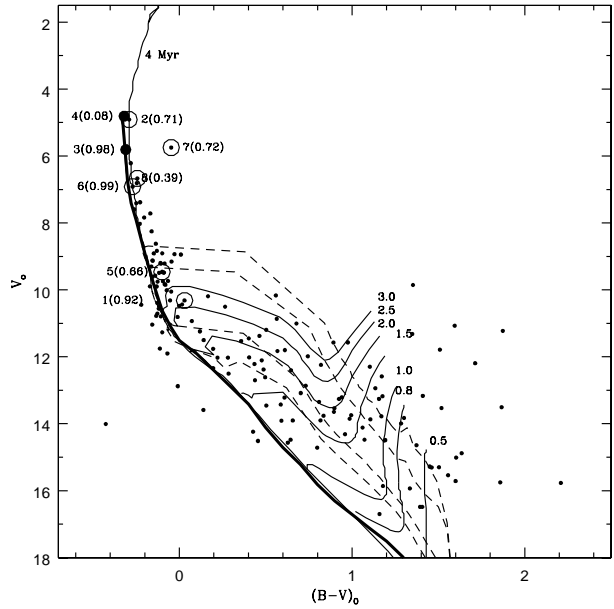}
\caption{Statistically cleaned $V_0,(B-V)_0$ CMD for stars lying in the cluster region.
The ZAMS (thick line) by  Schmidt-Kaler (1982), isochrone (thin continuous line) for 4 Myr age by Bertelli et al. (1984) 
and PMS isochrones of 0.5,1,5,10 Myr (dashed line) along with evolutionary tracks of 
different mass stars by Siess et al. (2000) are also shown. 
The MS and all the isochrones are corrected for a distance of 1 kpc. 
The stars shown by large open and filled circles have proper motion data. 
The star numbers are taken from WEBDA. The numbers in parentheses represent membership 
probability. The data for filled circles are taken from WEBDA.}
\end{figure*}

\begin{figure*}
\includegraphics[height=8cm,width=8cm]{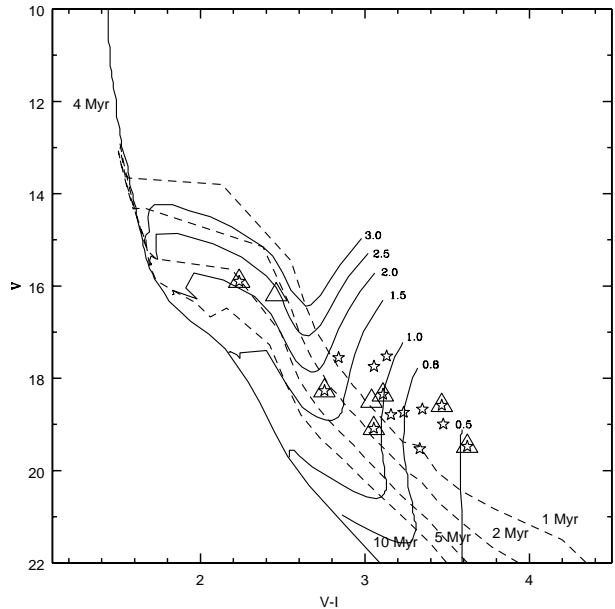}
\caption{ $V,(V-I)$ CMD for $H\alpha$ emission stars (star symbols) and NIR excess stars (triangles). 
Isochrone for 4 Myr by Bertelli et al. (1994, continuous line) and PMS isochrones for 1,2,5,10 Myr (dashed line)
along with evolutionary tracks of different mass stars by Siess et al. (2000) are also shown. 
All the isochrones are corrected for 
distance of 1 kpc and reddening $E(V-I)=1.75$ mag which corresponds to $E(B-V)=1.4$ mag. }
\end{figure*}

\begin{figure*}
\centering
\hbox{
\includegraphics[height=8.2cm,width=9cm]{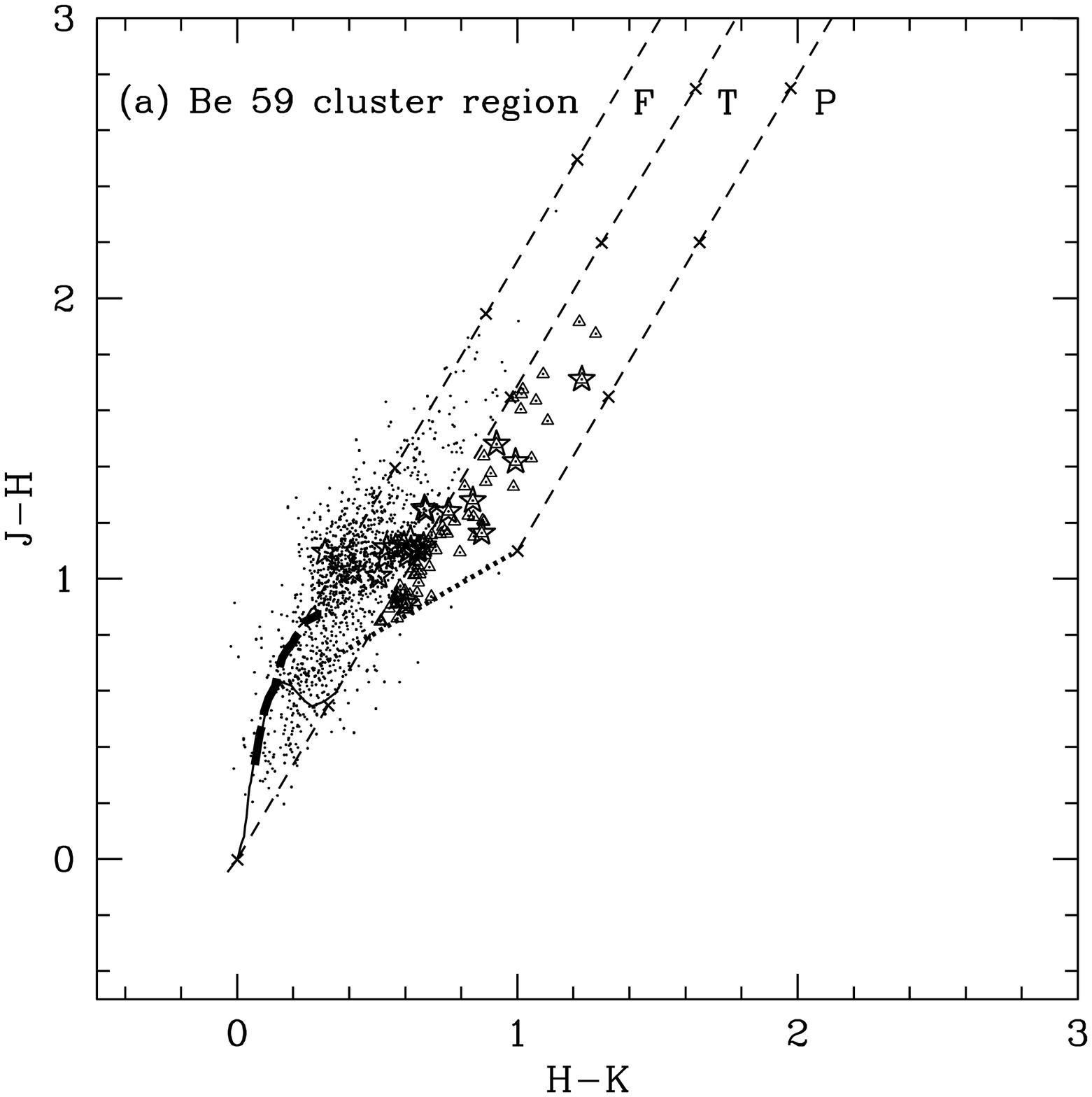}
\includegraphics[height=8.2cm,width=9cm]{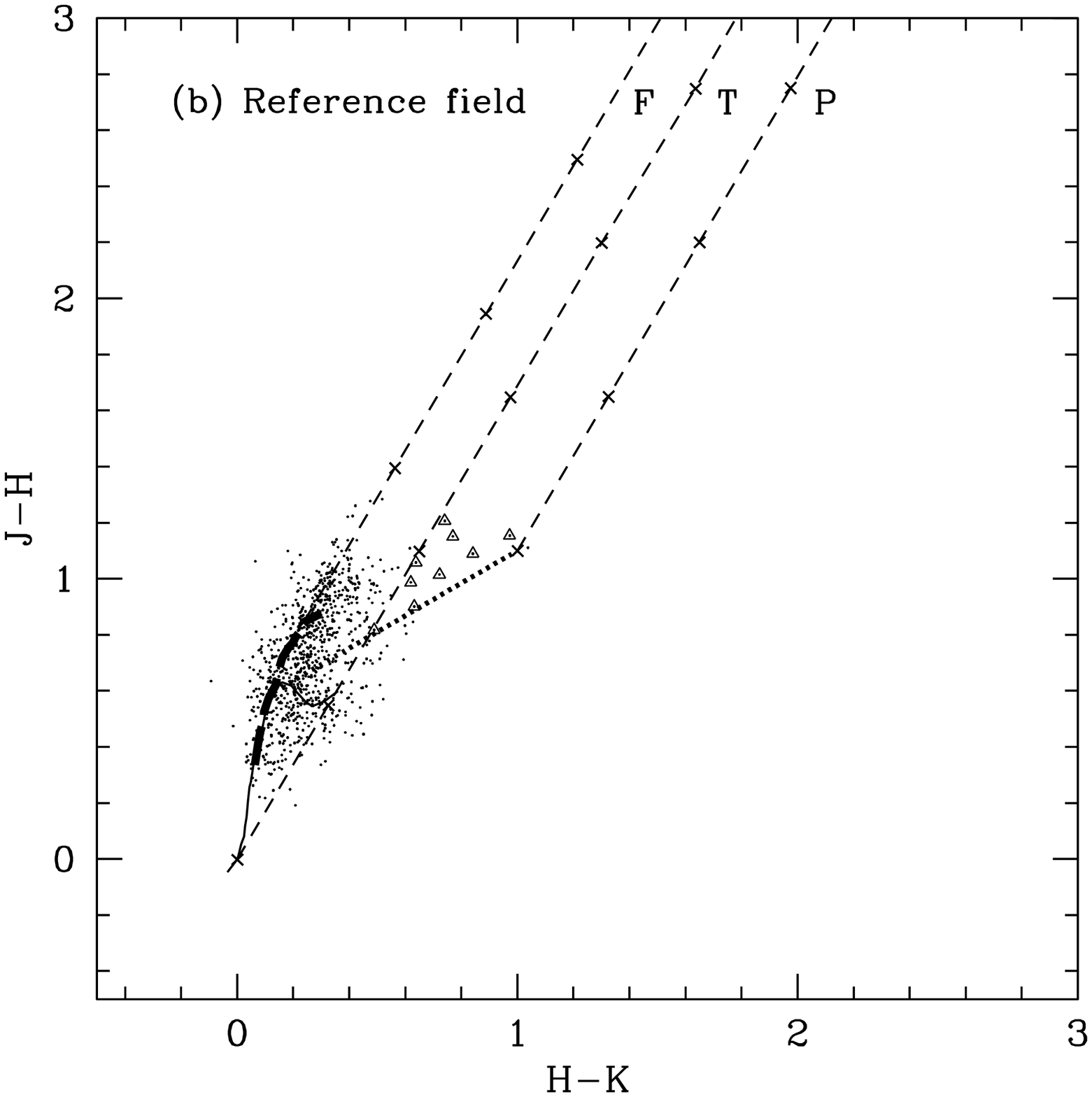}
}
\caption{$(J-H)/(H-K)$ colour-colour diagrams of sources detected in the $JHK_s$ bands 
with photometric errors less than 0.1 mag in (a) the cluster region ($r\le r_{cl}$) and 
(b) the reference field. The sequences for dwarfs (solid curve) and giants (thick dashed curve) 
are from Bessell \& Brett (1988). The dotted line represents the locus of T Tauri stars 
(Meyer et al. 1997). Dashed straight lines represent the reddening vectors (Cohen et al. 1981). 
The crosses on the dashed lines are separated by $A_V$ = 5 mag. The star symbols are the 
$H\alpha$ emission stars detected from slitless spectroscopy. Open triangles are probable 
T-Tauri stars.}

\end{figure*}

\begin{figure*}
\centering
\hbox{
\includegraphics[height=8.2cm,width=9cm]{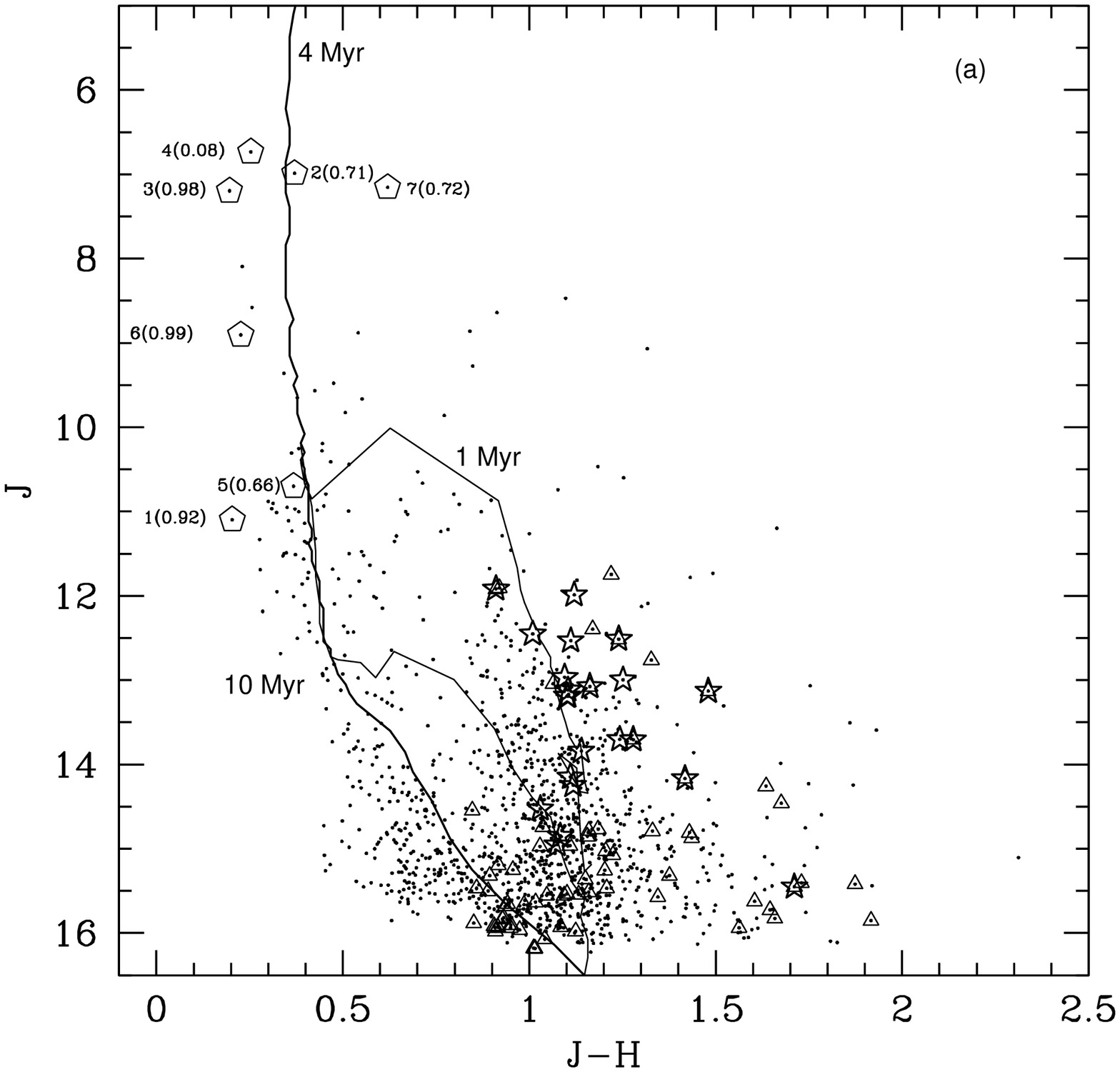}
\includegraphics[height=8.2cm,width=9cm]{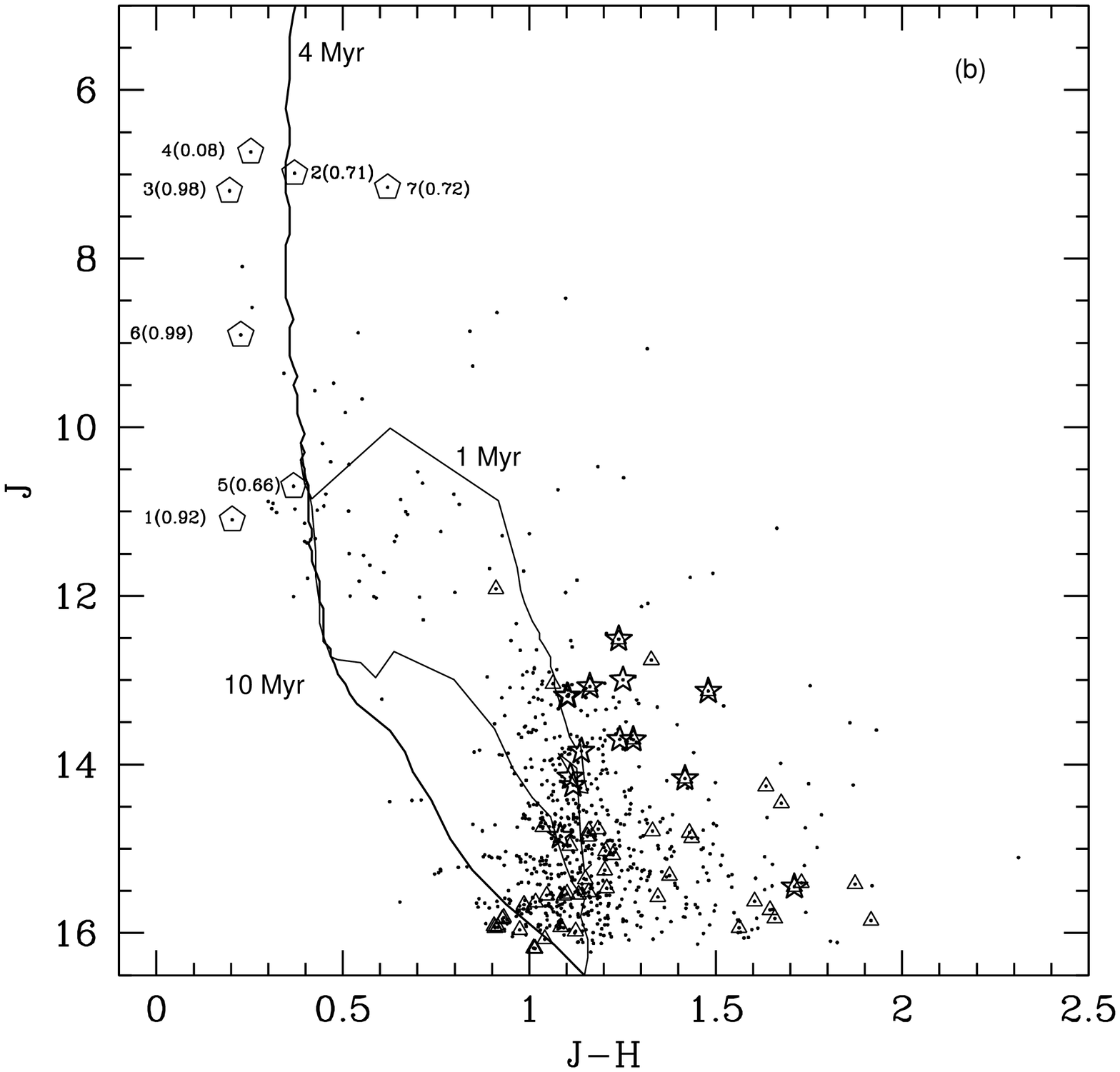}
}
\caption{(a) $J,(J-H)$ CMD of the stars within cluster region ($r\le r_{cl}$).
The symbols are same as in Fig. 11. The isochrone of 4 Myr (Z=0.02) by Bertelli et al. (1984) and PMS isochrones of age 1 and 10  Myr by Siess et al. (2000) corrected for a distance of 1 kpc and the minimum reddening $E(B-V)_{min}=1.4$ mag are also shown. The star numbers are taken from WEBDA. The numbers in parentheses represent membership probability.
(b) Same as (a) but statistically cleaned $J,(J-H)$ CMD. }
\end{figure*}

\begin{figure*}
\centering
\includegraphics[height=8cm,width=8cm]{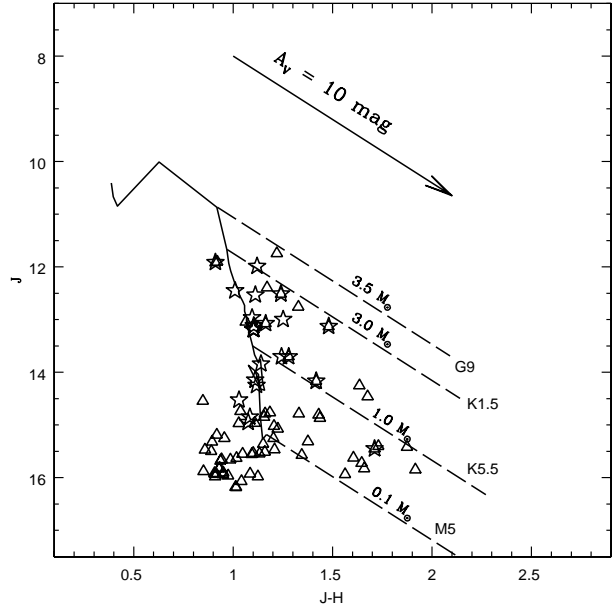}
\caption{$J,(J-H)$ CMD for the YSO candidates in cluster region ($r\le r_{cl}$).
The symbols are same as in Fig. 11. The solid curve denotes PMS isochrone of 1 Myr,
derived from Siess et al. (2000) and corrected for a distance of 1 kpc.
Masses range from 3.5 to 0.1 $M_\odot$
from top to bottom. The dashed oblique reddening lines are for PMS
stars of 0.1, 1, 3.0 and 3.5 $M_\odot$ from bottom to top. }
\end{figure*}

\begin{figure*}
\centering
\includegraphics[height=8cm,width=4cm,angle=-90]{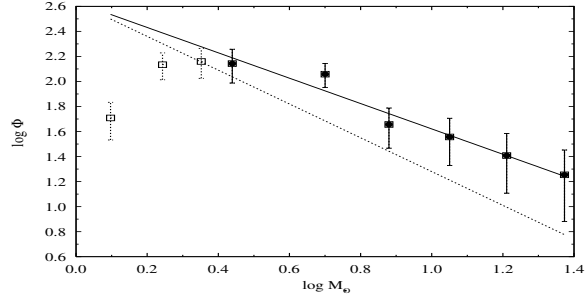}
\caption{A plot of the mass function in the cluster. log $\phi$ represents log ($N$/dlog $m$). The error bars represent $\pm\sqrt N$ errors. The continuous line shows a least square fit to the mass range $2.5<M/M_\odot <28$.
The dashed line represents mass function having Salpeter value $\Gamma = -1.35$.
 }
\end{figure*}

\begin{figure*}
\centering
\includegraphics[height=8cm,width=4cm,angle=-90]{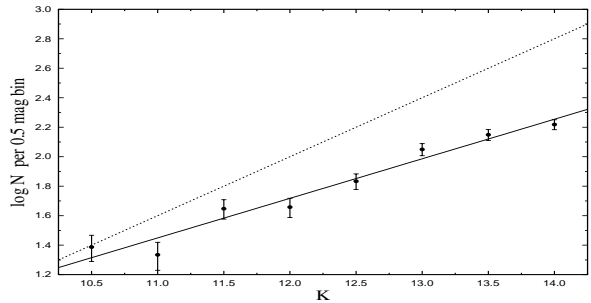}
\caption{ The corrected KLF for the probable members in the cluster. The star counts are number of stars per 0.5 mag bin and the error bars represents $\pm \sqrt N$ errors. The continuous line is the least-square fit to the data points in the mag range 10.5-14.0. The value of the power-law slope $\alpha$ is $0.27\pm0.02$. Dashed line having slope $\alpha =0.4$ represents the average KLF for young clusters (see text). }
\end{figure*}

\clearpage

\begin{figure*}
\centering
\includegraphics[height=14cm,width=16cm]{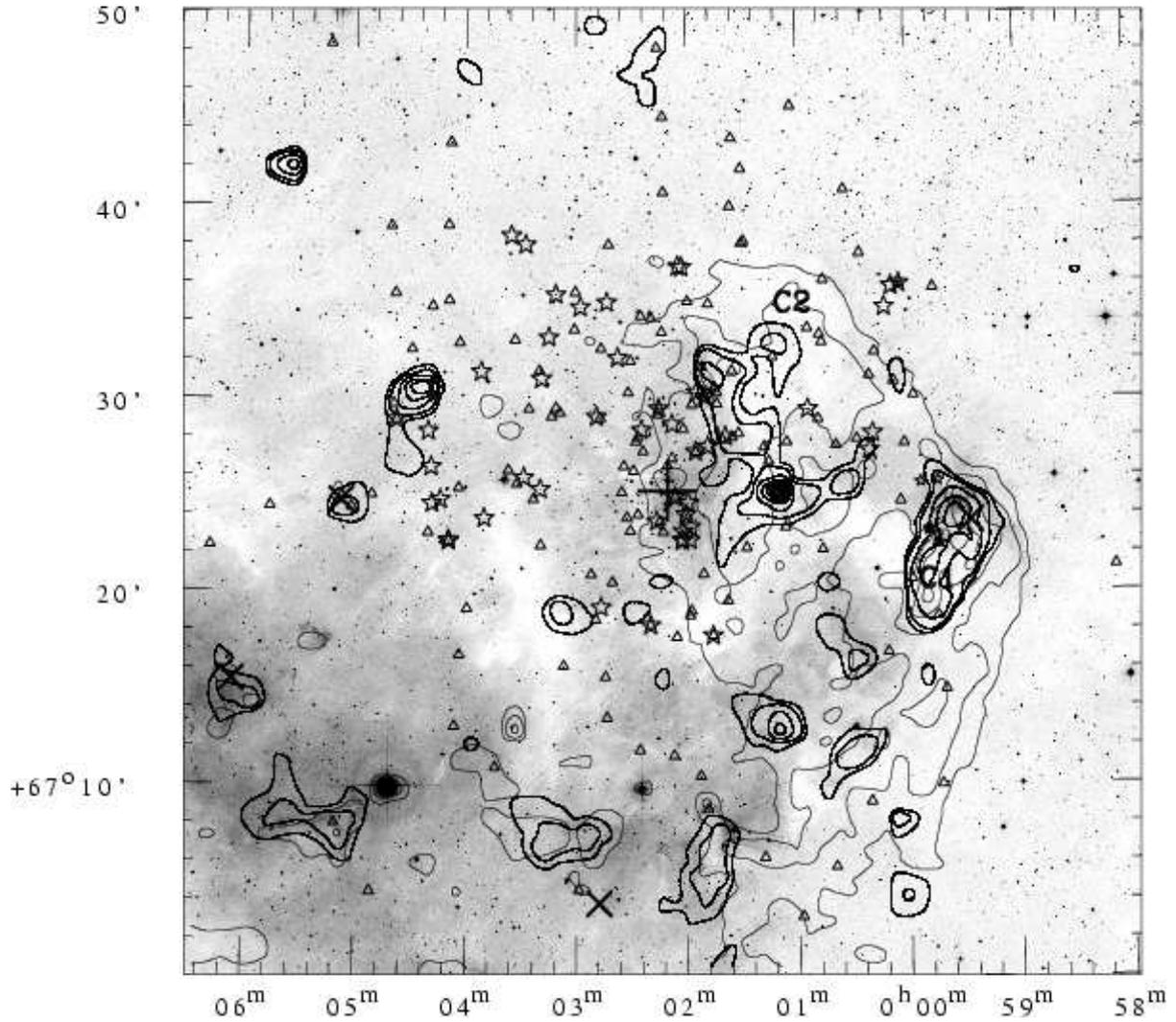}
\caption{Spatial distribution of IR-excess sources (open triangles), $H\alpha$ (star symbols) and  IRAS point sources (crosses) are overlaid on DSS-II $R$ band image. NVSS ($1.4~GHz$) radio contours (${\it thick ~contours}$) 
and MSX A-band intensity contours (${\it thin ~contours}$) have also been shown. 
The contours are 2, 4, 10, 20, 30, 40, 50, 60, 70, 80, 90 \% of the
peak value 134.6 mJy/beam for NVSS and 5, 10, 20, 30, 40, 50, 60, 70, 80, 90 \% of the
peak value $2.9 \times 10^{-5}$ $W m^{-2} Sr^{-1}$  for MSX A-band contours, respectively.
The abscissa and the ordinates are for the J2000 epoch.
C1 and C2 represent the locations of the peaks of molecular clumps (see Yang \& Fukui (1992) for details). The 
plus mark represents the optical center of the cluster.
}
\end{figure*}

\begin{figure*}
\centering
\hbox{
\includegraphics[height=6.5cm,width=8cm]{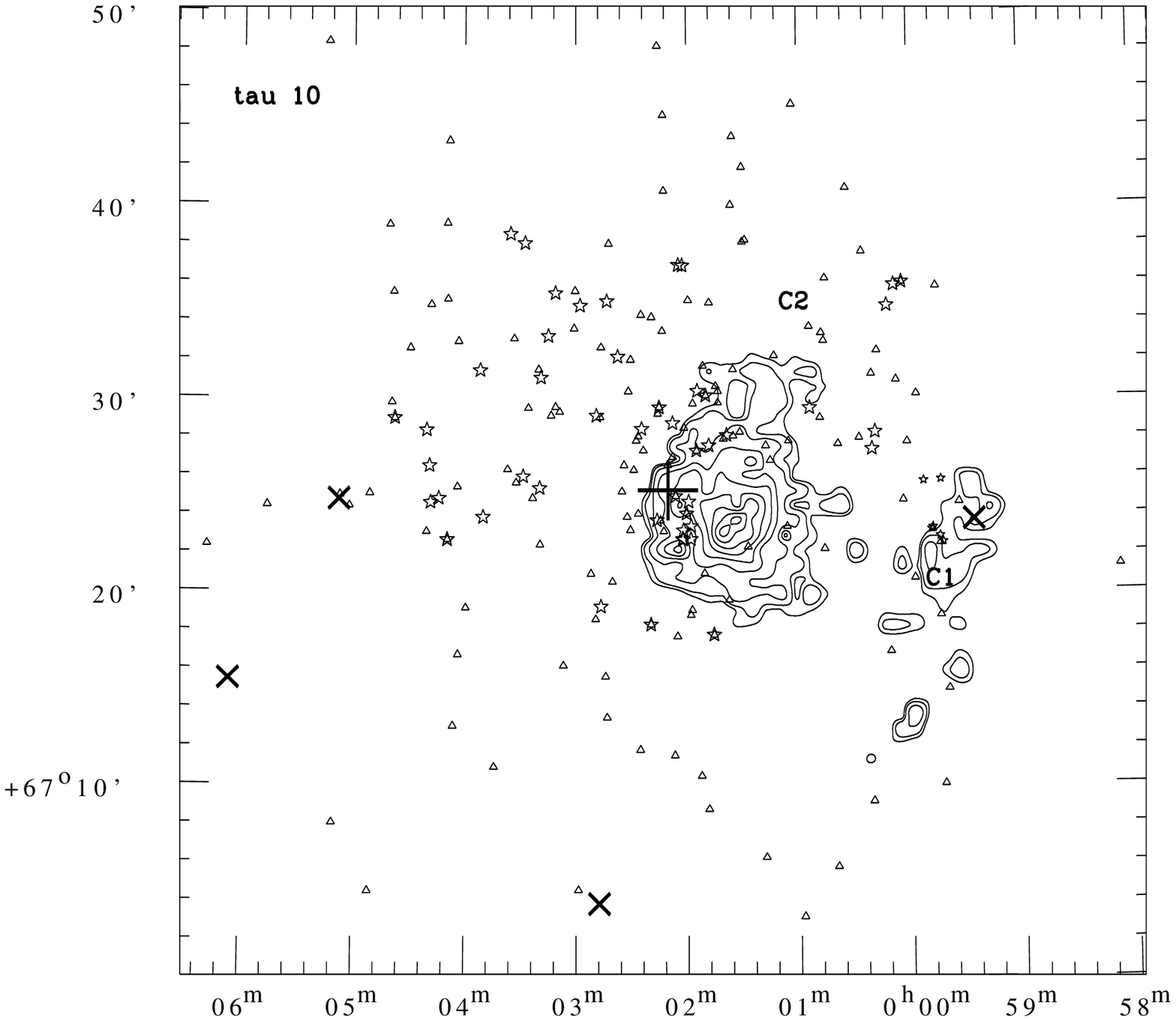}
\includegraphics[height=6.5cm,width=8cm]{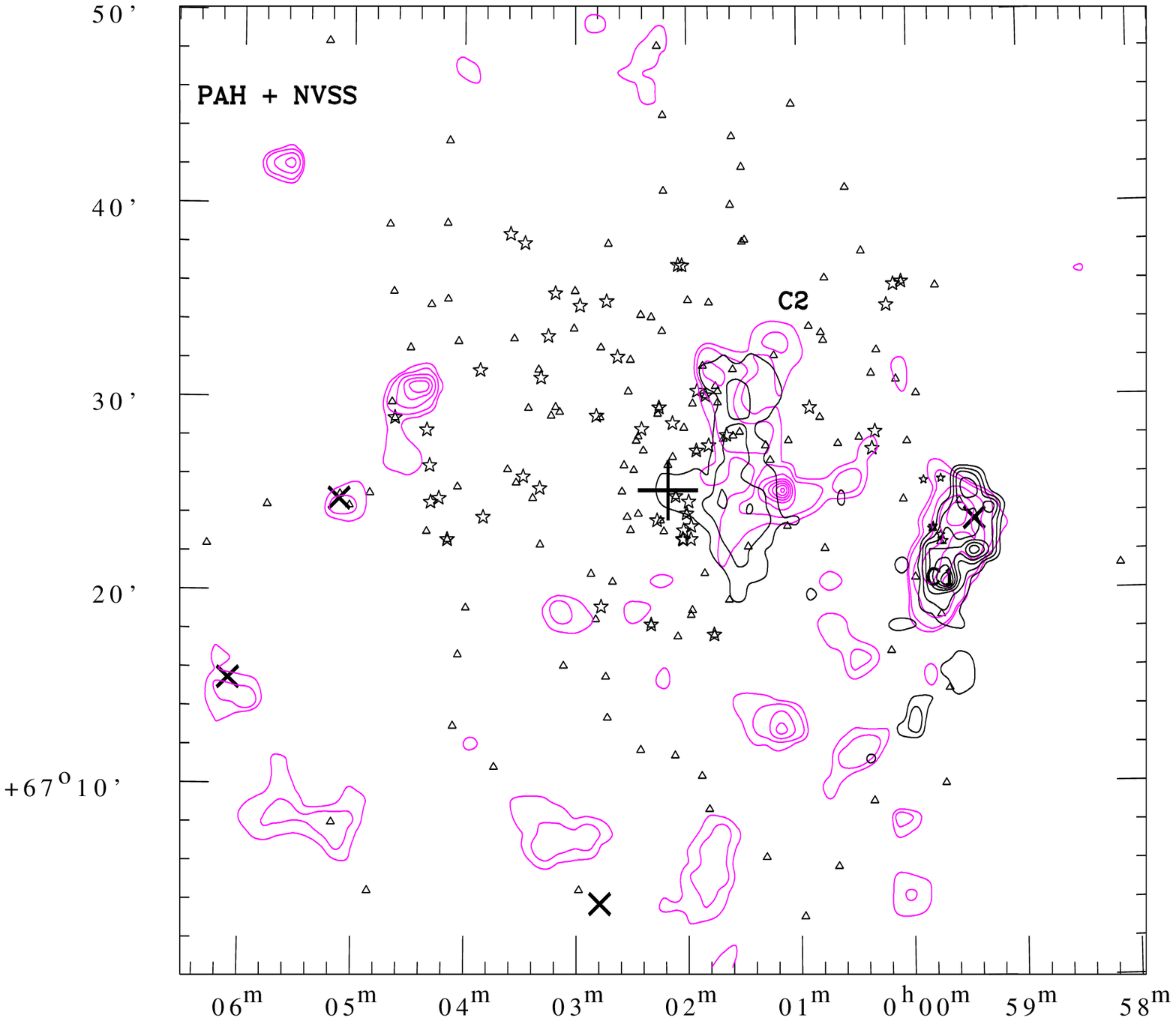}
}
\caption{ {\it(left)} Distribution of optical depth {($\tau_{10}$ at 10 $\mu$m)} 
around the cluster region. The contours are at 3, 5, 10, 20, 30, 40, 50, 60, 70, 80, 90 \% of 
the peak value of  $5.5\times 10^{-5}$.
{\it(right)} Emission in UIBs (thick contours) and radio continuum ($1.4~GHz$, thin contours). For UIBs the contours are at 10, 20, 30, 40, 50, 60, 70, 80, 90 \% of the peak value of $1.2\times10^{-5}$ $W m^{-2} Sr{^{-1}}$. The contour levels for radio continuum and the symbols are the same as in Fig. 16. The abscissa and the ordinates are in the J2000 epoch.}

\end{figure*}

\begin{figure*}
\centering
\hbox{
\includegraphics[height=6.5cm,width=8cm]{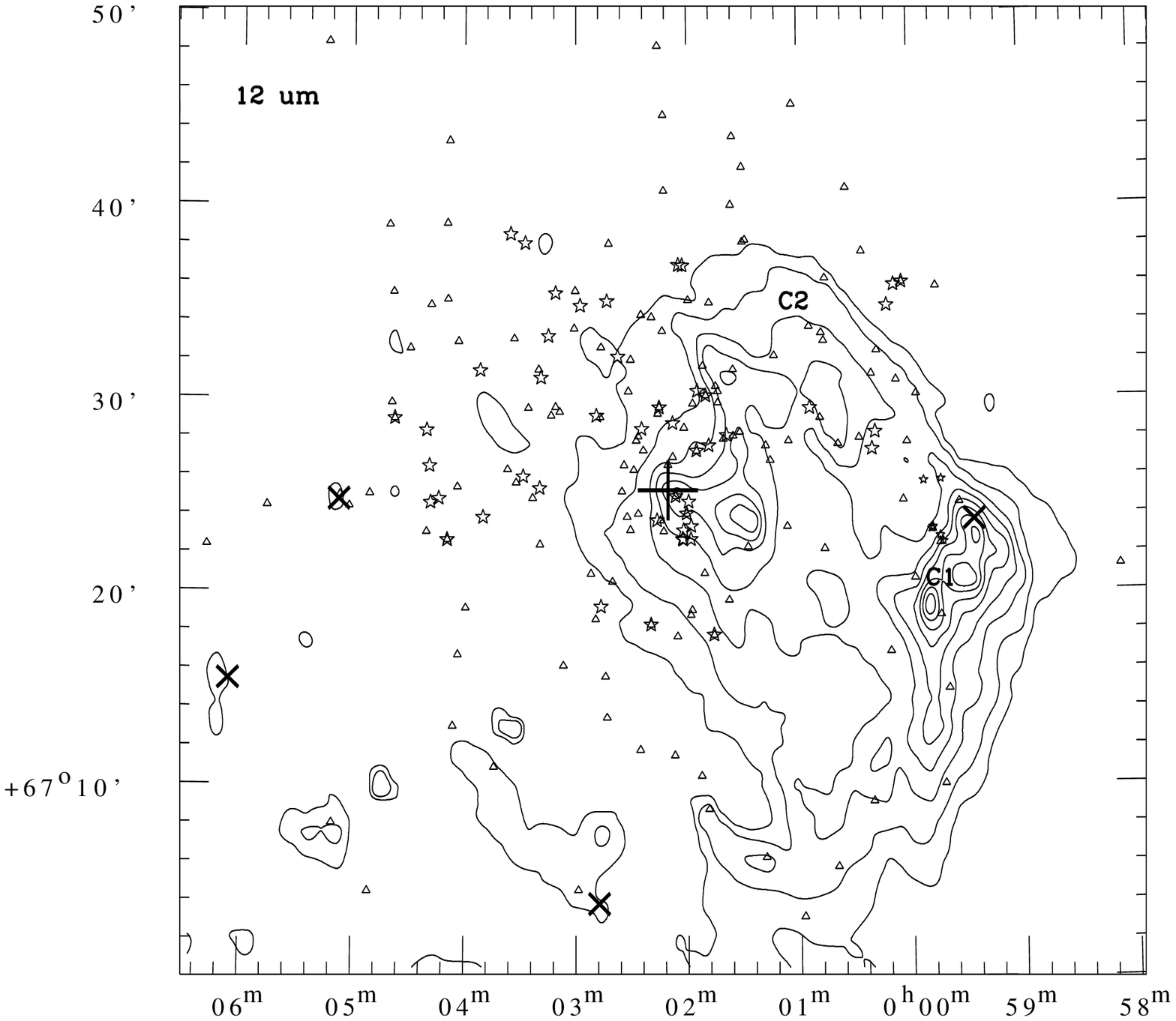}
\includegraphics[height=6.5cm,width=8cm]{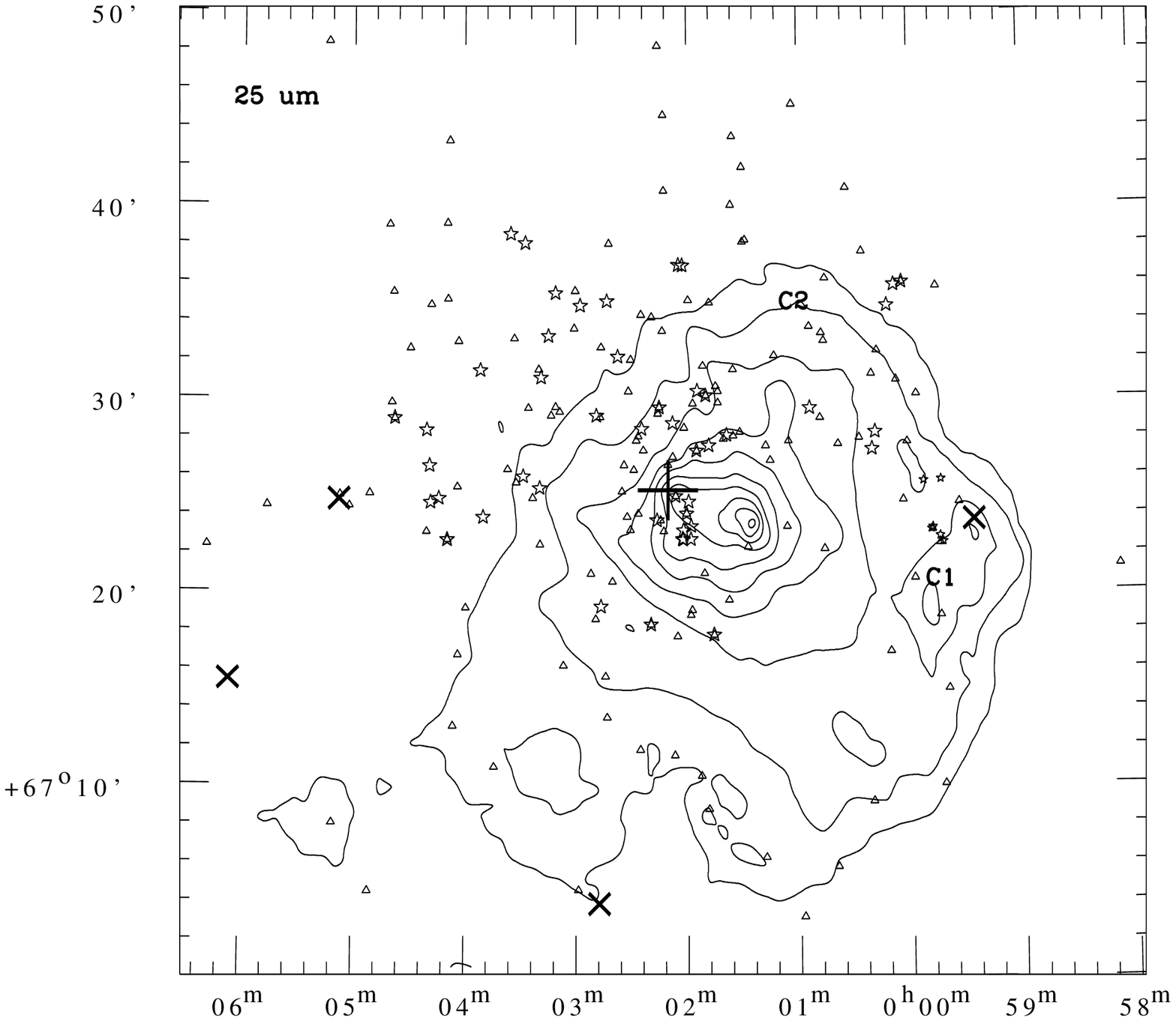}
}
\hbox{
\includegraphics[height=6.5cm,width=8cm]{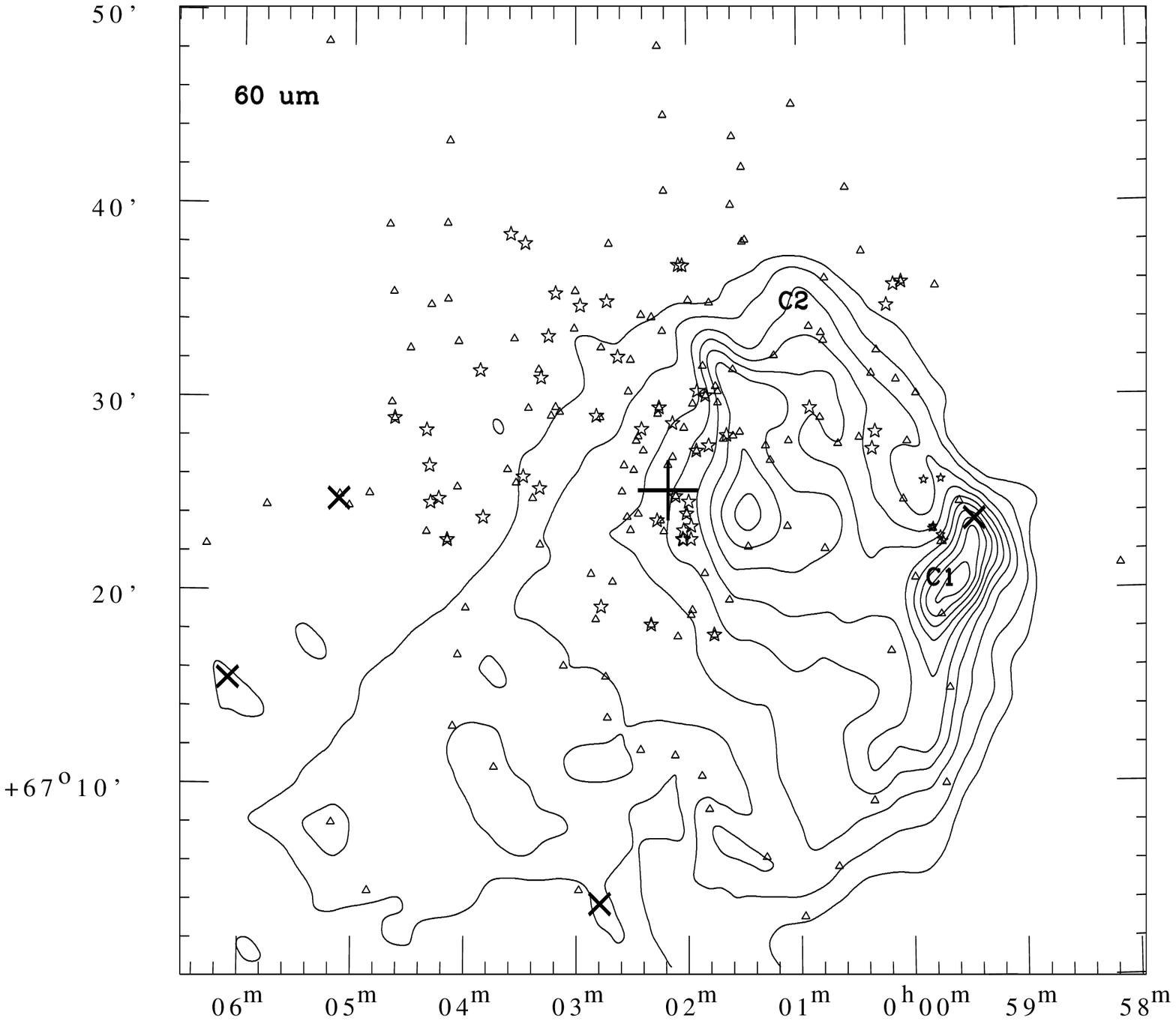}
\includegraphics[height=6.5cm,width=8cm]{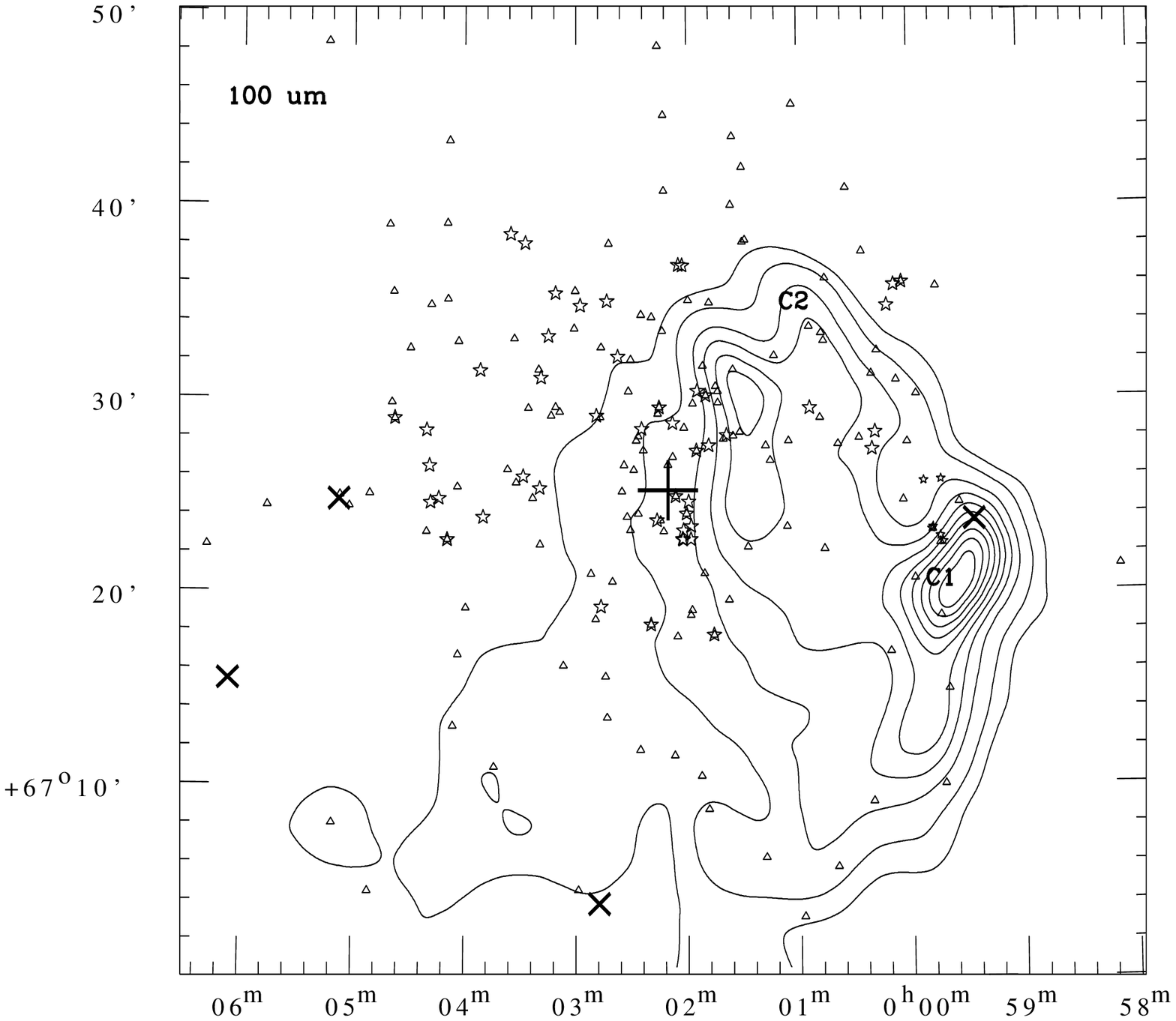}
}
\caption{The IRAS-HIRES intensity maps for the cluster region at $12~\mu m~ (top~ left)$, $25~ \mu m~ (top~ right)$,
$60~ \mu m~ (bottom~ left)$ and $100~ \mu m~ (bottom~ right)$.
The contours are at 5, 10, 20, 30, 40, 50, 60, 70, 80 and 90 \% of the peak value of 116 MJy/ster, 445 MJy/ster,
1560 MJy/ster and 2715 MJy/ster at 12, 25, 60 and 100 $\mu m$ respectively.
The symbols are as same as in Fig. 16.}
\end{figure*}

%\clearpage

\begin{figure*}
\centering
\hbox{
\includegraphics[height=6.5cm,width=8cm]{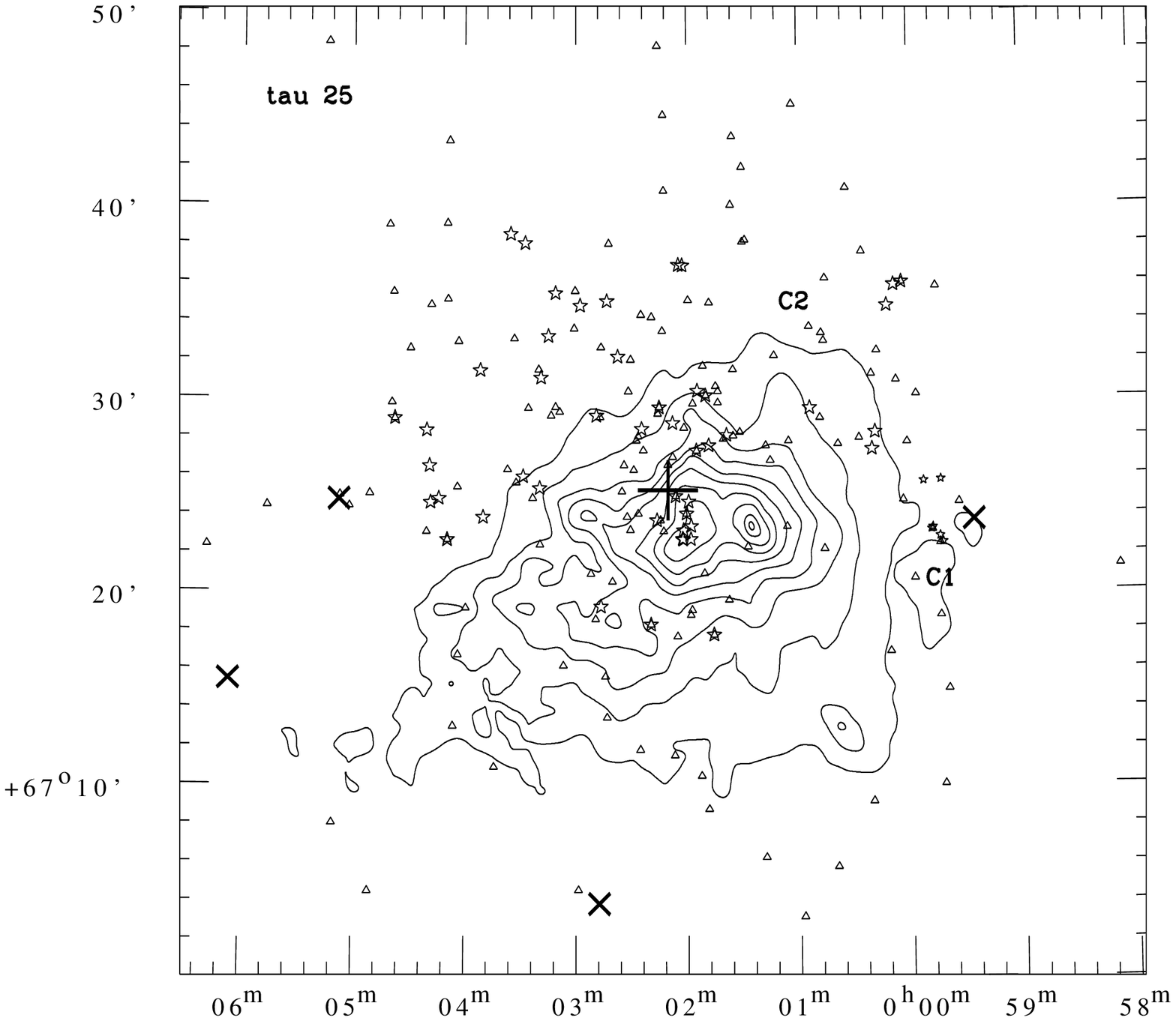}
\includegraphics[height=6.5cm,width=8cm]{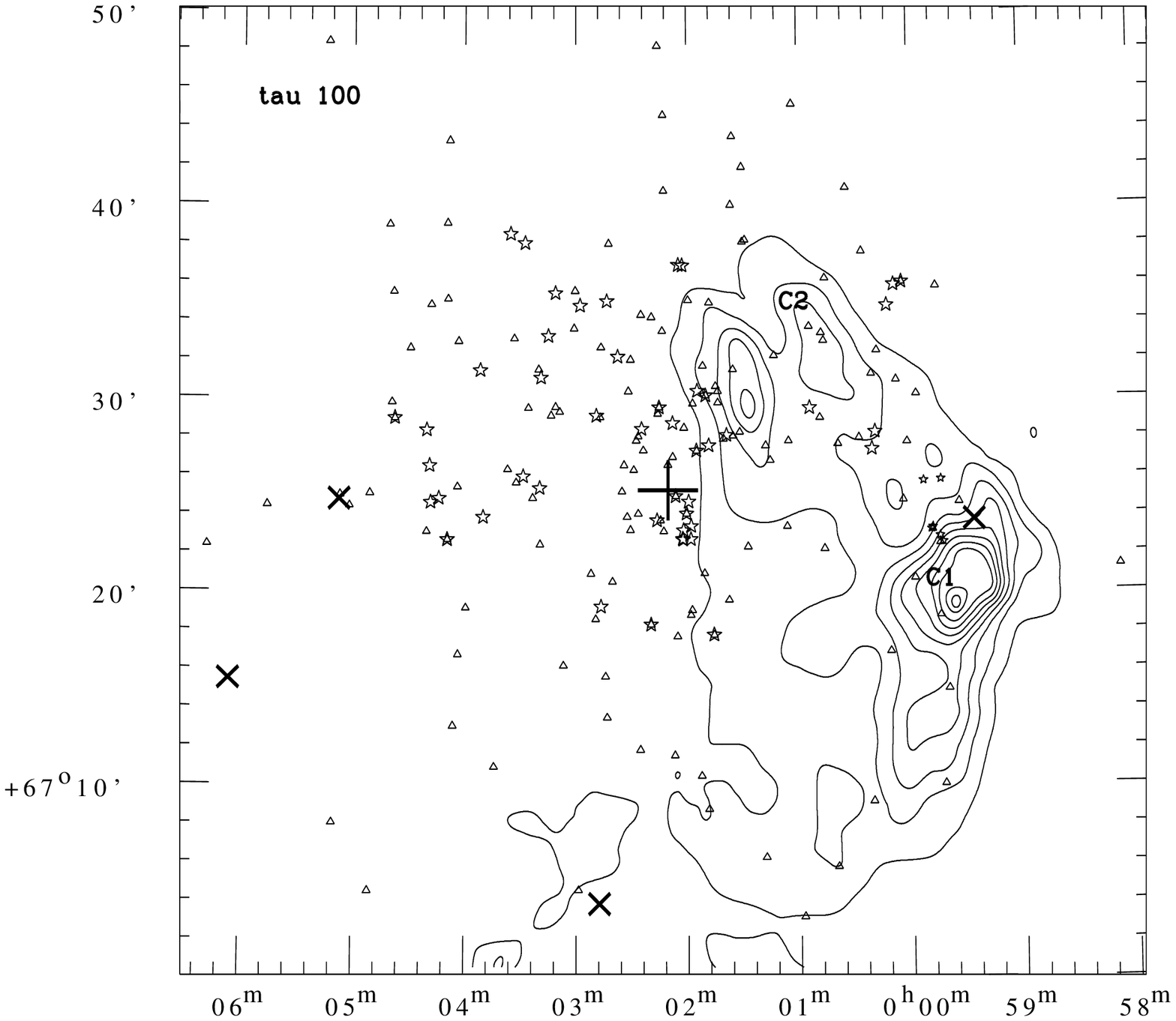}
}
\caption{
{\it Left panel}: The dust optical depth ($\tau_{25}$) distribution from the HIRES 12 and 25 $\mu m$ maps assuming 
a dust emissivity law of $\epsilon_\lambda \propto \lambda^{-1} $.
The contours represent 10, 20, 30, 40, 50, 60, 70, 80, 90, 95\% of the peak value of $1.09 \times  10^{-5}$.
{\it Right panel}: The dust optical depth ($\tau_{100}$) distribution from the HIRES 60 and 100 $\mu m$ maps. 
The contours represent 10, 20, 30, 40, 50, 60, 70, 80, 90, 95\% of the peak value of $3.83 \times 10^{-3}$.
The abscissa and the ordinates are in J2000 epoch.}

\end{figure*}

\begin{figure*}
\centering
\includegraphics[height=8cm,width=14cm]{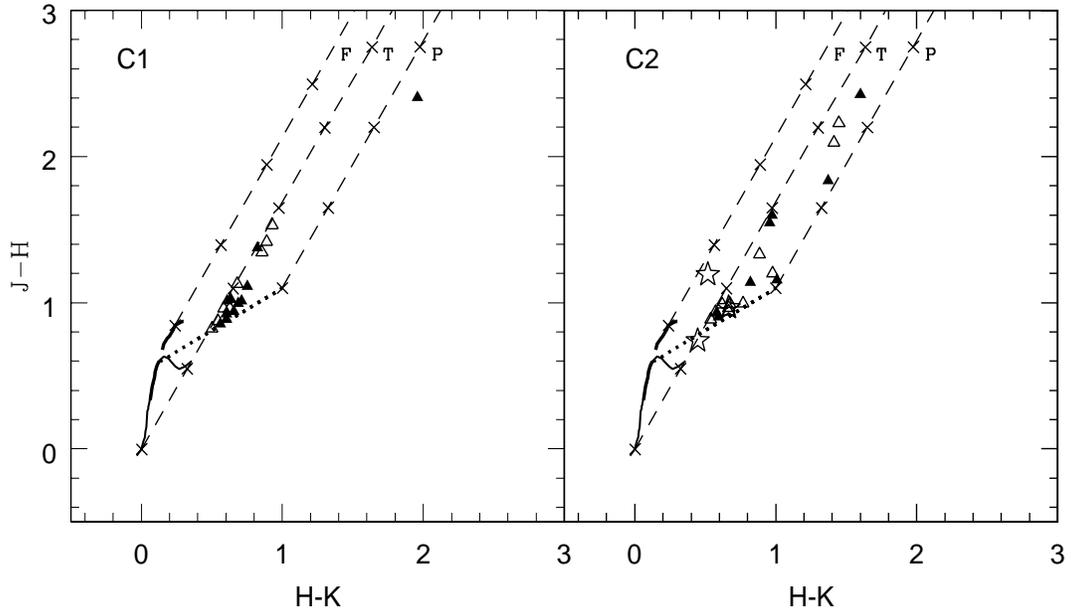}
\caption{$(J-H)/(H-K)$ CMD for the YSO candidates around the clumps C1 and C2.
The star symbols and triangles are H$\alpha$ emission and NIR excess stars, respectively. Filled triangles are stars having photometric errors between 0.1 to 0.2 mag.}
\end{figure*}

\begin{figure*}
\includegraphics[height=8cm,width=10cm]{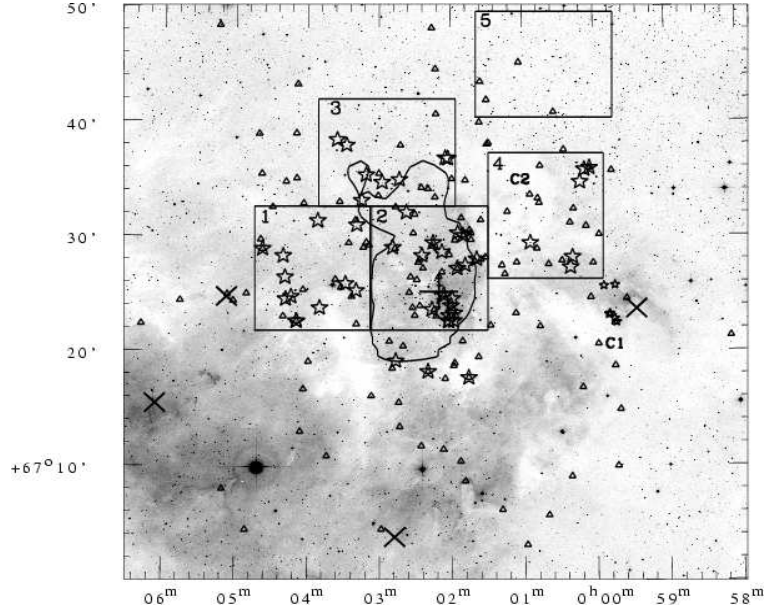}
\caption{Five selected sub-regions overlaid on DSS-II $R$ band image.
The symbols are the same as in Fig. 16. The abscissa and the ordinates are for the J2000 epoch.
The continuous curve represents the outer most 2MASS $K_s$-band isodensity contour (See Fig. 2 ({\it right panel})). }
\end{figure*}

\begin{figure*}
\includegraphics[height=4cm,width=14cm]{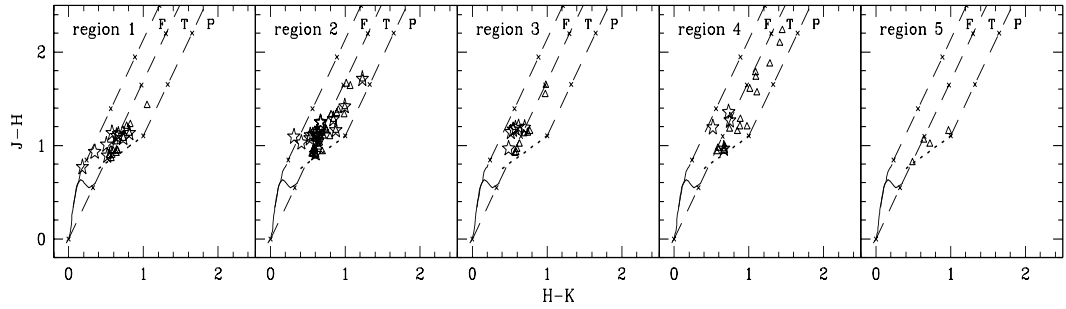}
\caption{$(J-H)/(H-K)$ CMD for the YSO candidates in 5 selected regions.
The symbols are same as in Fig. 11. }
\end{figure*}

\begin{figure*}
\includegraphics[height=6cm,width=14cm]{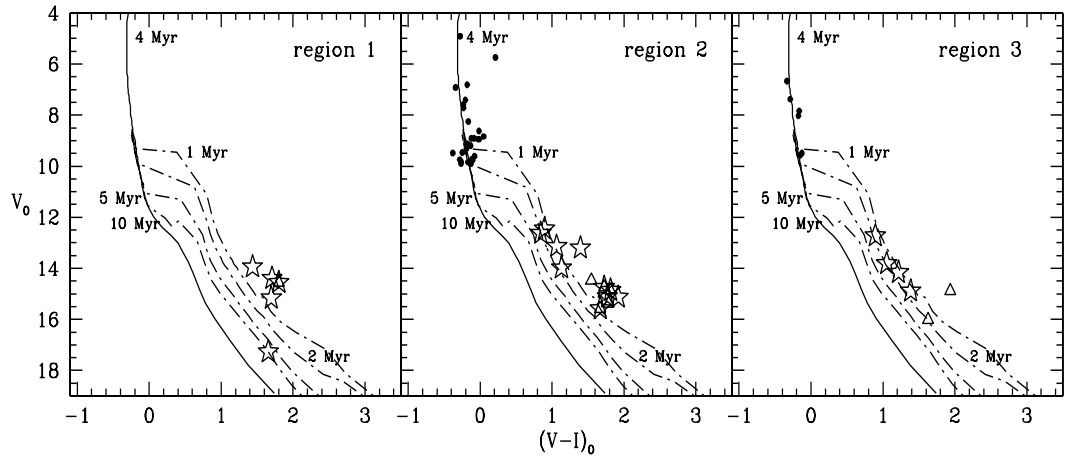}
\caption{$V_0/(V-I)_0$ CMD for probable YSOs lying in 3 selected regions.
The isochrone for 4 Myr age by Bertelli et al. (1994) and PMS isochrones
of 1,2,5,10 Myr by Siess et al. (2000) are also shown. All the isochrones are corrected for a distance of 1 kpc. The filled circles, triangles and star symbols represent MS, NIR excess and H$\alpha$ emission stars respectively.}
\end{figure*}

\begin{figure*}
\includegraphics[height=13cm,width=14cm]{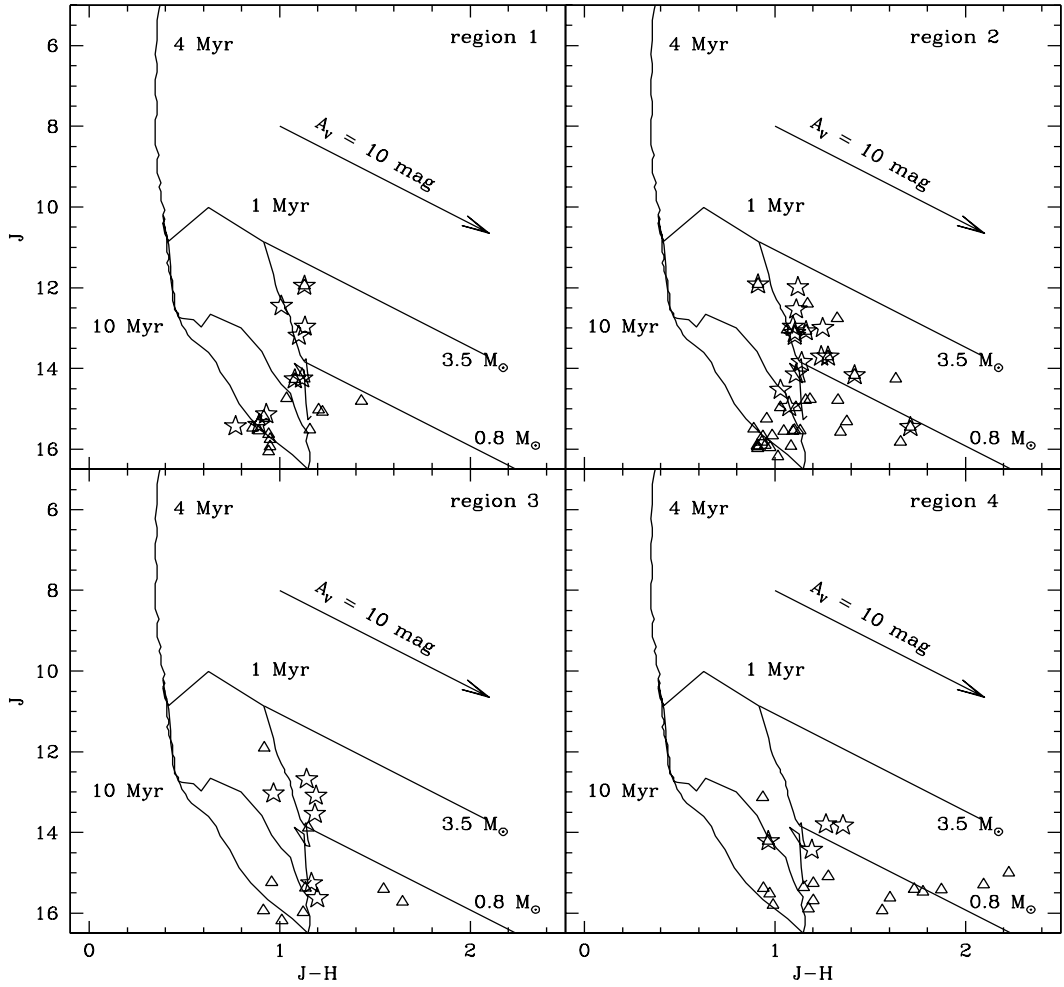}
\caption{$J,(J-H)$ CMD for probable YSOs lying in 4 selected regions. The symbols are the same as in Fig. 11. The isochrone of 4 Myr (Z=0.02)  by Bertelli et al. (1994) and PMS
isochrones of age 1 and 10 Myr by Siess et al. (2000) corrected for a distance of 1 kpc 
and reddening $E(B-V)_{min}=1.4$ mag are also shown. The continuous oblique lines denote the reddening vectors for PMS stars of 0.8 and 3.5 $M_\odot$. The straight line indicates reddening vector for A$_V$= 10 mag. }
\end{figure*}

\begin{figure*}
\includegraphics[height=8cm,width=4cm,angle=-90]{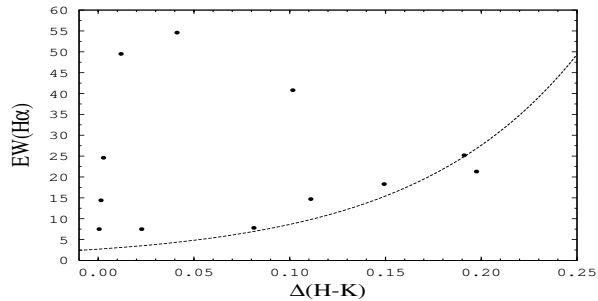}
\caption{$H\alpha$ equivalent width EW($H\alpha$) in ${\rm \AA}$ vs NIR-excess ($\Delta (H-K)$) diagram. The dash curve represent eye-estimated lower profile for the data points shown in figure 10 of Herbig \& Dahm (2002). }

\end{figure*}

\begin{figure*}
\includegraphics[height=4cm,width=8cm]{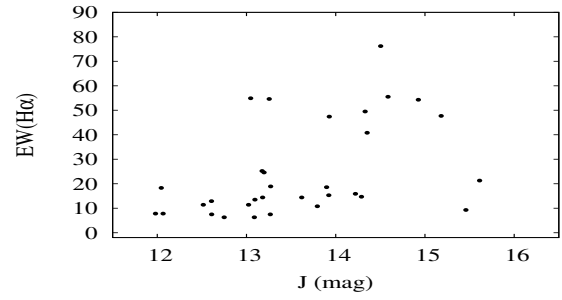}
\caption{ $H\alpha$ equivalent width, EW($H\alpha$), in ${\rm \AA}$ as a function of $J$ mag. }
\end{figure*}

\label{lastpage}

\end{document}